\documentclass[letterpaper]{article}
\usepackage{chemformula} 
\usepackage[T1]{fontenc} 
\usepackage{amssymb}
\usepackage{amsmath}
\usepackage{geometry}
\usepackage{setspace}
\usepackage{enumitem}
\usepackage{graphicx}
\usepackage{subcaption}
\usepackage{float}
\usepackage{authblk}
\usepackage[version = 4]{mhchem}
\usepackage[style = chem-acs, doi = true, articletitle=true]{biblatex}

\newfloat{scheme}{htbp}{los}
\floatname{scheme}{Scheme}
\floatname{chart}{Chart}
\newfloat{graph}{htbp}{loh}

\author[1]{Lan Hoang Mai}
\author[2]{Nazifa Tasnim Arony}
\author[2]{R S Joshya}
\author[1]{Tyler Manfre}
\author[2]{Joshua M. O. Zide}
\author[1, 2]{Matthew F. Doty}
\affil[1]{Quantum Science and Engineering Program, University of Delaware, Newark, Delaware 19716, United States}
\affil[2]{Department of Materials Science and Engineering, University of Delaware, Newark, Delaware 19716, United States}

\title{A high-Q Split Cavity Enabling Independent Electrical Tuning of two Quantum Dots}

\date{*Email: lanmai@udel.edu, doty@udel.edu}

\begin{document}

\maketitle

\begin{abstract}
Cavity-mediated strong coupling is an essential tool for quantum photonics. However, the inherent inhomogeneity in ensembles of matter-based qubits necessitates independent tuning to bring more than one qubit into resonance with a single cavity mode. We present the design, fabrication, and characterization of a GaAs-based photonic crystal split cavity that consists of two electrically-distinct portions, which preserves the ability to apply independent electric fields that would independently tune two distinct qubits based on InAs Quantum Dots (QDs) into resonance with a cavity mode. We show that this device, despite the cut required to electrically isolate the two halves, achieves an average Q $\geq 20,000$, sufficient to enter the strong coupling regime. Moreover, we show that the Q is limited primarily by the precision and accuracy of the e-beam lithography tool, rather than by sidewall scattering. This result creates a realistic experimental path toward devices that use controlled interactions between two InAs QDs, mediated by strong coupling to a single cavity mode, to generate multi-dimensional photon graph states.
\end{abstract}

\section*{Keywords}
InAs QDs, ultra-thin diode, split-cavity, independent tunability, cavity QED

\section*{Abbreviations}
InAs, QDs, PhCC, QED

\section{Introduction}
A quantum dot-cavity system provides important functionality for photonic quantum information processing. For example, quantum dots (QDs) serve as bright sources of on-demand, indistinguishable single photons \cite{Uppu_scalable_2020, Nowak_deterministic_2014, Liu_bright_2013, Kolatschek_bright_2021, Hauser_deterministic_2026, Somaschi_near-optimal_2016, Zeuner_on-demand_2021}, hosts for spin qubits \cite{Economou_scalable_2012, Prechtel_decoupling_2016, Warburton_single_2013, Appel_coherent_2021, Bechtold_quantum_2016, Delley_deterministic_2017}, and photon cluster state generators for measurement-based quantum computing \cite{Cogan_deterministic_2023, Vezvaee_deterministic_2022, Su_continuous_2024, Istrati_sequential_2020, Huet_deterministic_2025, Russo_photonic_2018}. Coupling such QDs to photonic cavities can enhance the photon emission rate (Purcell effect, weak coupling regime) or lead to vacuum Rabi splitting (strong coupling regime). Of particular interest is the opportunity to couple two QD-based spin qubits to a single photonic cavity mode, allowing strong coupling to mediate qubit-qubit interactions. Exploiting that strong coupling to create entanglement between the two qubits provides a mechanism for generating two-dimensional cluster states of polarization-encoded photonic qubits, which are a universal resource for measurement-based quantum computing \cite{Economou2008, Economou_optically_2010, Buterakos_deterministic_2017}. 

The scalable production of QD-cavity devices is severely limited by the spectral inhomogeneity of the QDs. For InAs QDs, this inhomogeneity results from the random diffusion inherent to Stranski–Krastanov growth. Several methods have been developed to overcome the spectral inhomogeneity of epitaxial quantum dots, including electrical tuning via the Stark effect \cite{Laucht2010, Nowak_deterministic_2014, Schall_bright_2021, Somaschi_near-optimal_2016}, magnetic tuning \cite{Peniakov_magneto-optics_2025, Kim_strong_2011}, temperature tuning \cite{Ding_on-demand_2016, Unsleber_highly_2016}, and strain tuning \cite{Yang_tunable_2024, Lettner_strain-controlled_2021}. Among these techniques, electric field tuning is most desirable due to its fast switching speeds, wide tuning range, and ease of experimental integration. For example, it is relatively easy to create diode structures that apply a uniform electric field to all QDs embedded within a sample. However, this does not allow the application of a unique electric field to each individual QD. To date, methods employed to independently tune individual QDs have included applying voltage exclusively to one of the stacked quantum dot layers \cite{Kim_independent_2009}, implanting protons to laterally isolate the diode structure \cite{Thon_independent_2011}, performing a shallow diode etch to electrically separate two quantum dots coupled to the same photonic crystal waveguide \cite{Chu_independent_2023, Hallett_controlling_2026}, and applying strain tuning to multiple quantum dots within a shared suspended beam waveguide \cite{Grim_scalable_2019}. 

We present the design, fabrication, characterization, and modeling of a split cavity consisting of two coupled L3 PhC cavities separated by a monolithic cut. We first use finite-difference time-domain (FDTD) simulations to optimize the split cavity design by adjusting the holes immediately adjacent to the L3 cavities while maintaining structural symmetries. We then fabricate the cavity and characterize its quality factor via the cross-polarized resonant scattering technique. We achieve an experimental quality factor (Q) as high as $6.2 \times 10^4$. Finally, we simulate the quality factor distribution under realistic fabrication disorders and show that the Q is limited primarily by the precision and accuracy of the e-beam lithography tool rather than by scattering loss. Together, these results show that our split cavity can enable localized, independent electrical tuning of two quantum dots while preserving the high quality factor required to achieve strong coupling between the quantum dots and the cavity.

\section{Split cavity design}
\begin{figure}[H]
    \centering
    \includegraphics[width=0.5\linewidth]{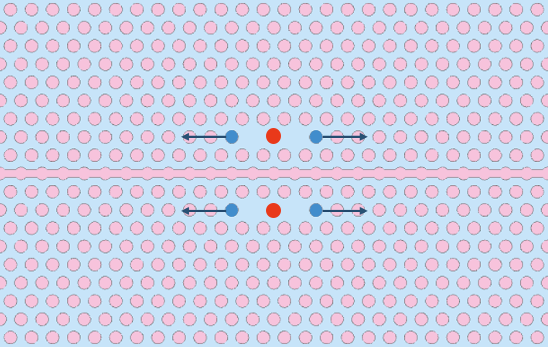}
    \caption{Schematic of the split cavity. Blue circles indicate end holes whose position and radius are computationally optimized. Red circles indicate the intended QD locations.}
    \label{fig:cavity}
\end{figure}

As shown in Fig.~\ref{fig:cavity}, the split cavity consists of two evanescently-coupled L3 cavities separated by a monolithic cut that creates two electrically-distinct diode structures (top and bottom). The photonic crystal has a nominal lattice constant $a = 249$ nm, a nominal hole radius $r = 0.294a \approx 73$ nm, and a slab thickness of $0.602a \approx 150$ nm \cite{Carfagno_computational_2023}. The cut is designed to be 90 nm wide, spanning the entire width of the split cavity. This specific width was chosen to minimize scattering loss while remaining above the minimum feature size of our fabrication process. We envision that site-templated QD growth methods \cite{McCabe_low-density_2020, McCabe_techniques_2021} would be used to position two QDs at the locations indicated by the red dots in Fig.~\ref{fig:cavity}, which are each approximately 431 nm away from the center of the cut. Previous COMSOL simulations by Carfagno et al.~\cite{Carfagno_computational_2023} demonstrate that the stray potential (cross-talk) for two QDs separated by 862 nm is less than $10 \mu$V, which is several orders of magnitude lower than the hundreds of millivolts typically required for spectral tuning \cite{Lobl_narrow_2017}. Thus the electric field applied to the top and bottom portions of the split cavity could be chosen to independently tune each QD into resonance with a cavity mode.

\begin{figure}[H]
    \centering
    \begin{subfigure}[htbp]{0.495\linewidth}
        \centering
        \includegraphics[width=\linewidth]{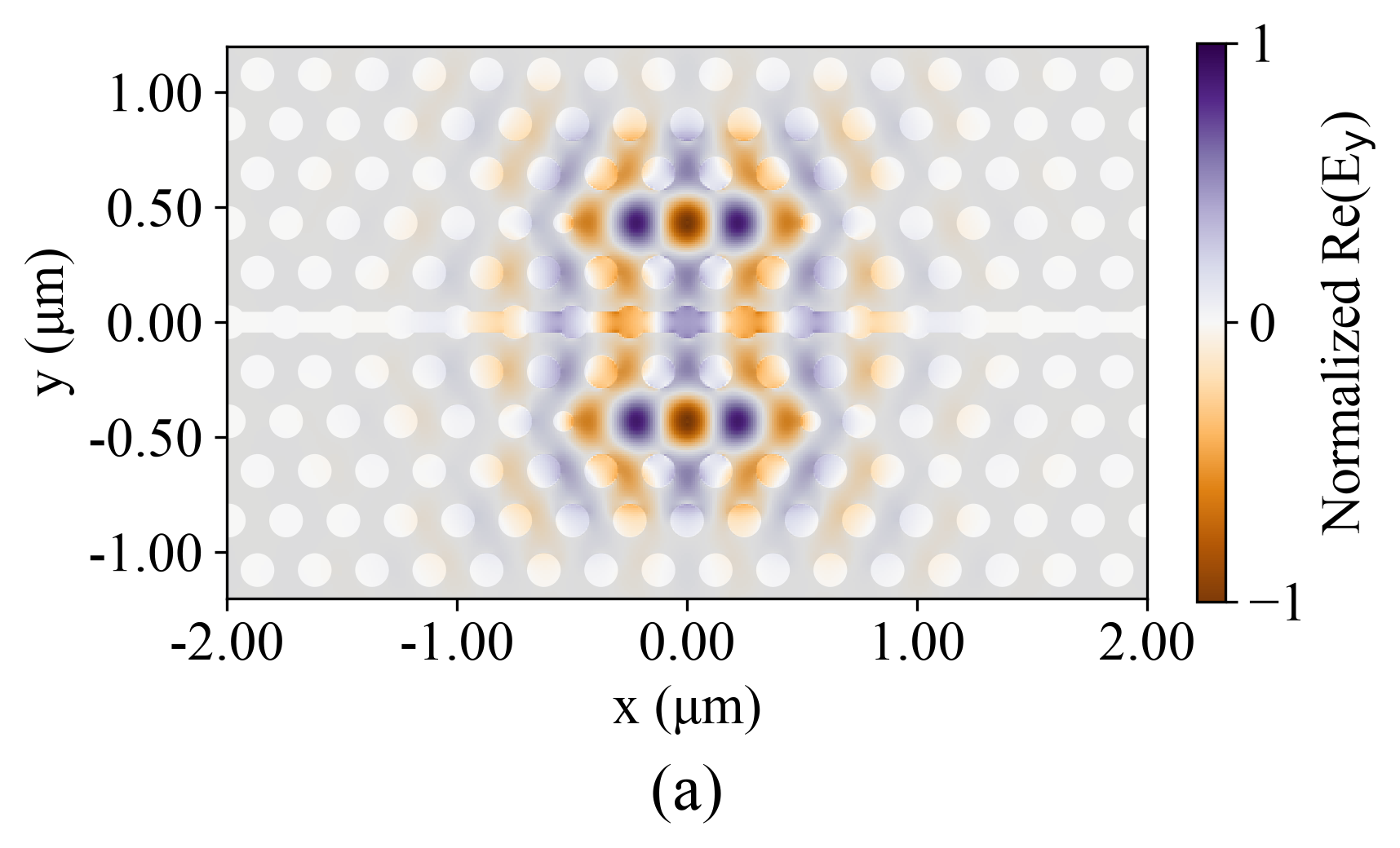}
        \label{fig:Ey_real_even}
    \end{subfigure}
    \hfill
    \begin{subfigure}[htbp]{0.495\linewidth}
        \centering
        \includegraphics[width=\linewidth]{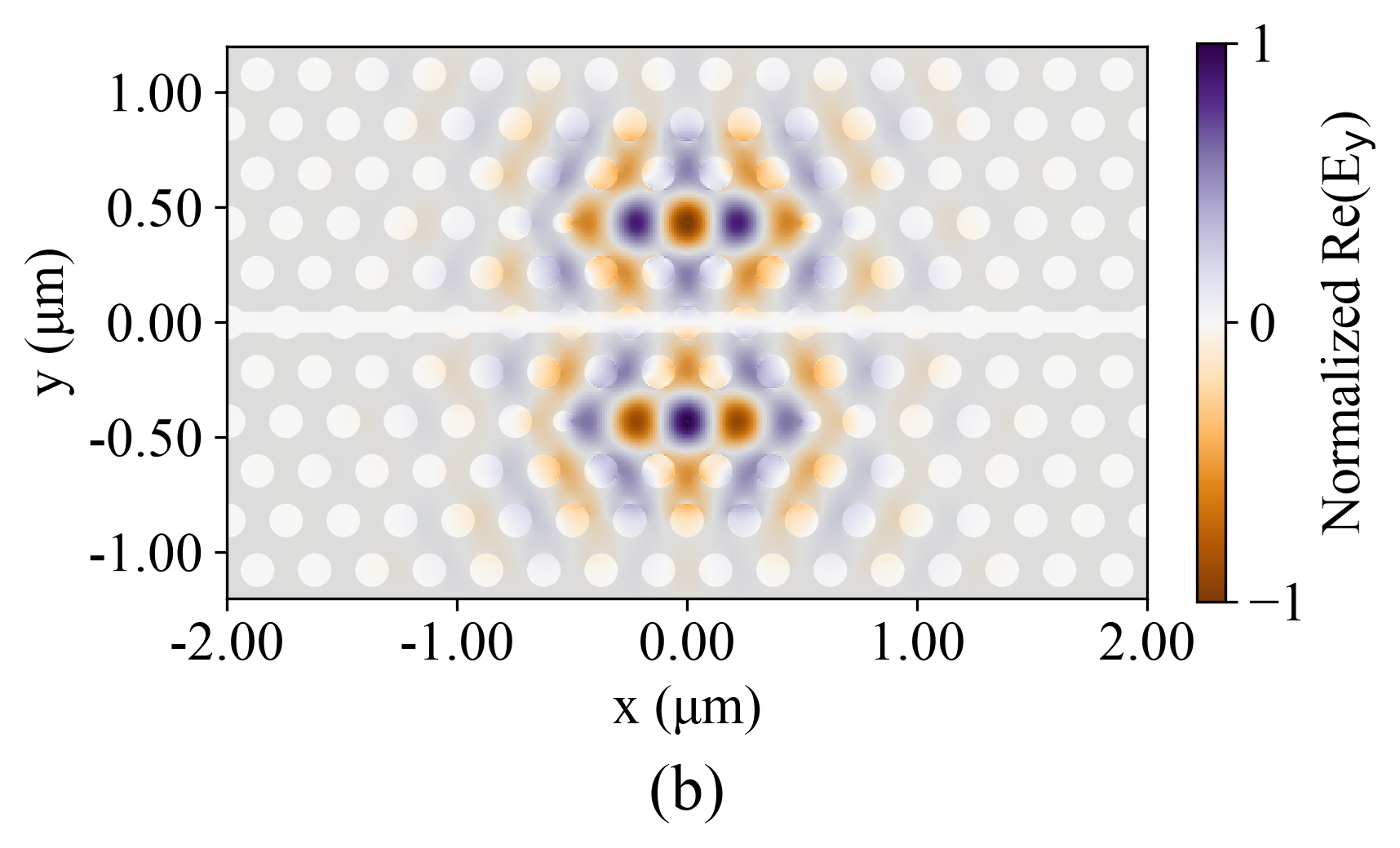}
        \label{fig:Ey_real_odd}
    \end{subfigure}
    \caption{(a) $\vec{E_y}$ field of the \textbf{even} cavity mode at resonant wavelength of $\lambda = 964.231$ nm. (b) $\vec{E_y}$ field of the \textbf{odd} cavity mode at resonant wavelength of $\lambda = 962.539$ nm.}
     \label{fig:Ey_real}
\end{figure}

Because it consists of two coupled L3 cavities, the split cavity exhibits two distinct modes: a symmetric (even) mode, where the electric field distribution in each L3 cavity shares the same parity, and an anti-symmetric (odd) mode, where the field distributions possess opposite parities. These modes are shown in Fig.~\ref{fig:Ey_real}a and b, respectively. The odd mode should ideally be used for \ce{InAs} QD-cavity coupling; it generally exhibits a higher Q than the even mode because the node in the electric field at the cut minimizes scattering loss. However, as discussed further below, the method we used to characterize the cavity Q is primarily sensitive to the even mode. Consequently, we computationally analyze both the even and odd modes, but fabricate and characterize only the device optimized for even mode performance.  

We first computationally optimize the Q of the split cavity by adjusting the horizontal position ($\Delta x$) and radius ($\Delta r$) of the end holes. These optimized end holes are denoted in blue in Fig.~\ref{fig:cavity}. To maintain the cavity's mirror symmetries, all end holes are simultaneously shifted in the directions indicated by the black arrows. For each selected value of ($\Delta x, \Delta r$), a FDTD simulation is performed to determine the cavity Q. In the simulation, the cavity resonance is excited by a dipole source centered at 945 nm with a bandwidth of 71 nm. To conform to the periodicity of the photonic crystal, the simulation mesh parallel to the photonic crystal plane is divided into grid sizes of $a/24$ and $(\sqrt{3}/2)(a/24)$ in the x and y directions, respectively. Perfectly matched layer boundary conditions are implemented to prevent spurious field reflections at the simulation boundaries. The total simulation runtime is set to 10 ps to capture sufficient cavity mode decay for an accurate Q-factor calculation.

\begin{figure}[H]
    \centering
    \includegraphics[height=0.9\textheight, keepaspectratio]{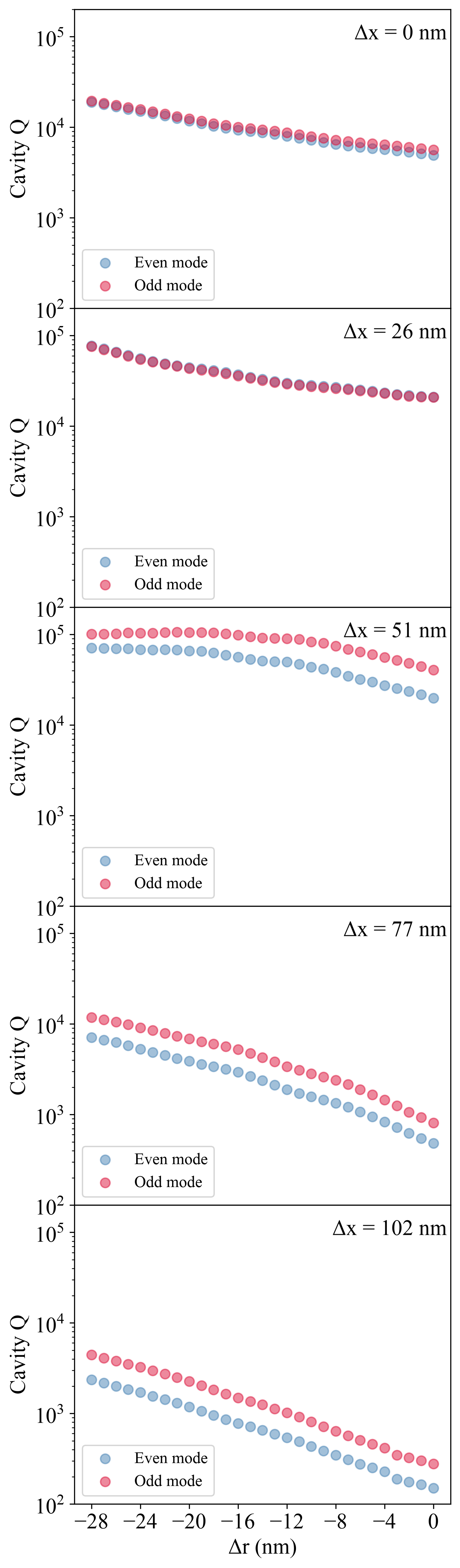}
    \caption{Sampling the optimization landscape with two parameters: radius shrink $\Delta r$ and horizontal shift $\Delta x$ of the end holes. The five panels show the cavity's quality factor (Q) for $\Delta r$ ranging from 0 to -28 nm for each of $\Delta x =$ 0 nm, 26 nm, 51 nm, 77 nm, 102 nm.}
    \label{fig:Q_delr}
\end{figure}

The cavity is optimized over a $\Delta x$ range from 0 nm to 102 nm and a $\Delta r$ range from -28 nm to 0 nm. These ranges are chosen to avoid overlapping adjacent air holes and to comply with the minimum fabricable diameter of 90 nm. Assuming a step size of 1 nm for both parameters, a full grid search over this parameter space would require 2,987 FDTD simulations. To reduce computational overhead, we first sample the parameter space by selecting five equally spaced $\Delta x$ values (0 nm, 26 nm, 51 nm, 77 nm, and 102 nm) and, for each $\Delta x$, compute the performance for $\Delta r$ values ranging from -28 nm to 0 nm in 1 nm steps. The resulting quality factors are plotted in Fig.~\ref{fig:Q_delr}. In all configurations except for $\Delta x = 51$ nm, the Q of both modes is maximized at the minimum end-hole radius ($\Delta r = -28$ nm). 

\begin{figure}[H]
    \centering
    \includegraphics[width=\linewidth]{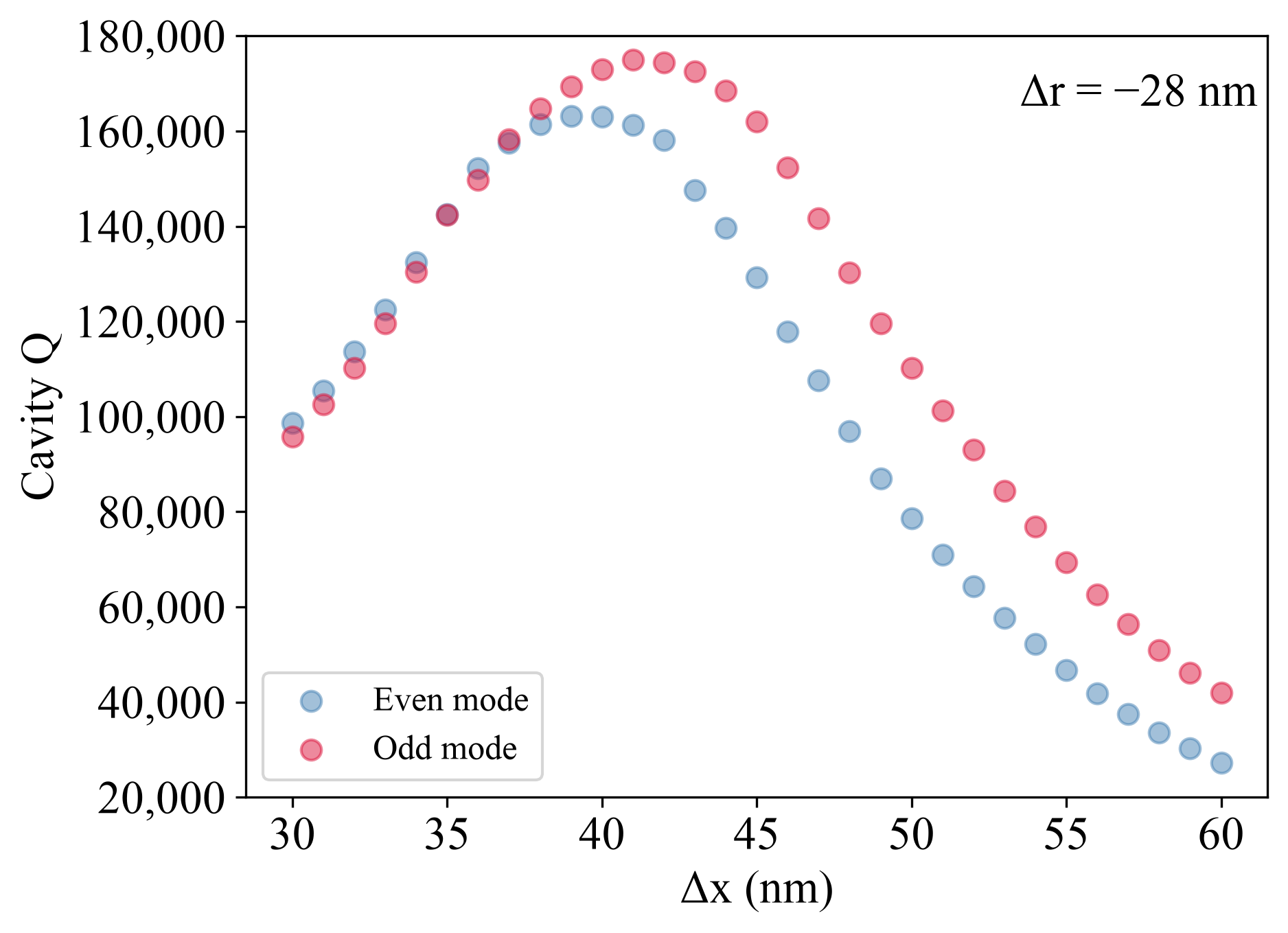}
    \caption{Cavity Q calculated for $\Delta r = -28$ nm and for $\Delta x = 30$ nm to $\Delta x = 60$ nm in 1 nm increments.}
    \label{fig:Q_delx_even_odd}
\end{figure}

For $\Delta x = 51$ nm, the dependence of Q on $\Delta r$ is qualitatively different than that for other values of $\Delta x$ and the maximum Q is substantially higher. This identifies the target parameter space for further optimization. Using a fixed value of $\Delta r = -28$ nm, we compute the Q as a function of end hole horizontal position for $\Delta x$ ranging from $\Delta x = 30$ nm to $\Delta x = 60$ nm in 1 nm increment. As shown in Fig.~\ref{fig:Q_delx_even_odd}, we find a maximum even mode Q of 163,113 at $\Delta x = 39$ nm and a maximum odd mode Q of 175,005 at $\Delta x = 41$ nm. These parameters define the design that we proceed to fabricate and characterize.

\section{Device Fabrication and Characterization}
\begin{figure}[H]
    \centering
    \includegraphics[width=\linewidth]{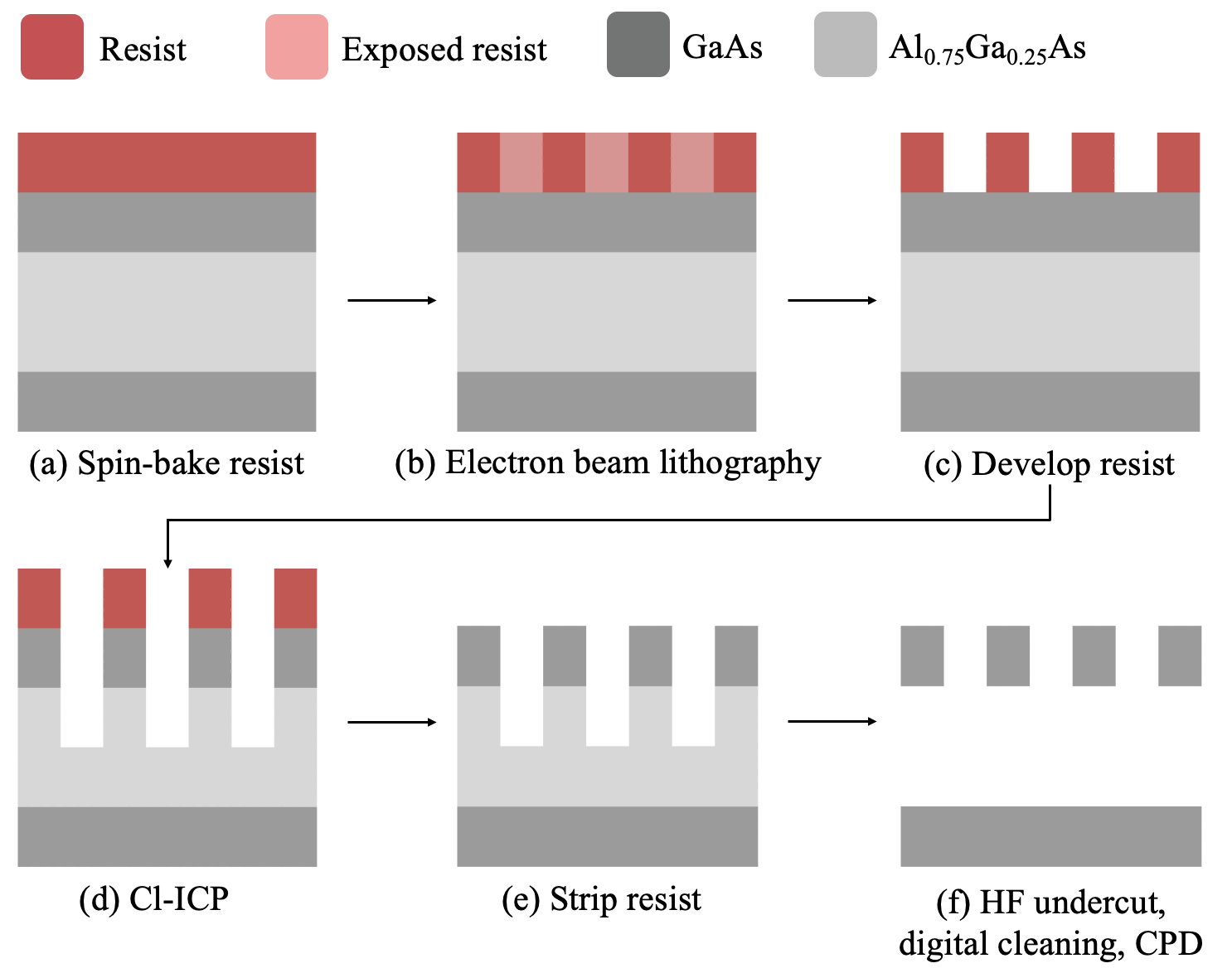}
    \caption{Fabrication process schematic. See SI for fabrication details.}
    \label{fig:fab_process}
\end{figure}

Fig.~\ref{fig:fab_process} shows the cross-sectional view of the main fabrication steps. The split cavity is fabricated on a p-i-p-i-n diode embedded within a 150-nm thick \ce{GaAs} membrane. This diode design is inspired by Löbl et al.~\cite{Lobl_narrow_2017}, who utilized a p-i-n-i-n diode structure for the single-electron charging of quantum dots. In our case we envision using \ce{InAs} QDs charged with a single hole \cite{Economou_scalable_2012, Vezvaee_deterministic_2022} so the middle diode layer is modified from n-type to p-type. The material stack is grown via molecular beam epitaxy on a (001) \ce{GaAs} substrate and the membrane is grown on top of a 1370-nm thick \ce{Al_{0.75}Ga_{0.25}As} sacrificial layer that is later etched to suspend the \ce{GaAs} membrane.  

\begin{figure}[H]
    \centering
    \begin{subfigure}[htbp]{0.495\linewidth}
        \centering
        \includegraphics[width=\linewidth]{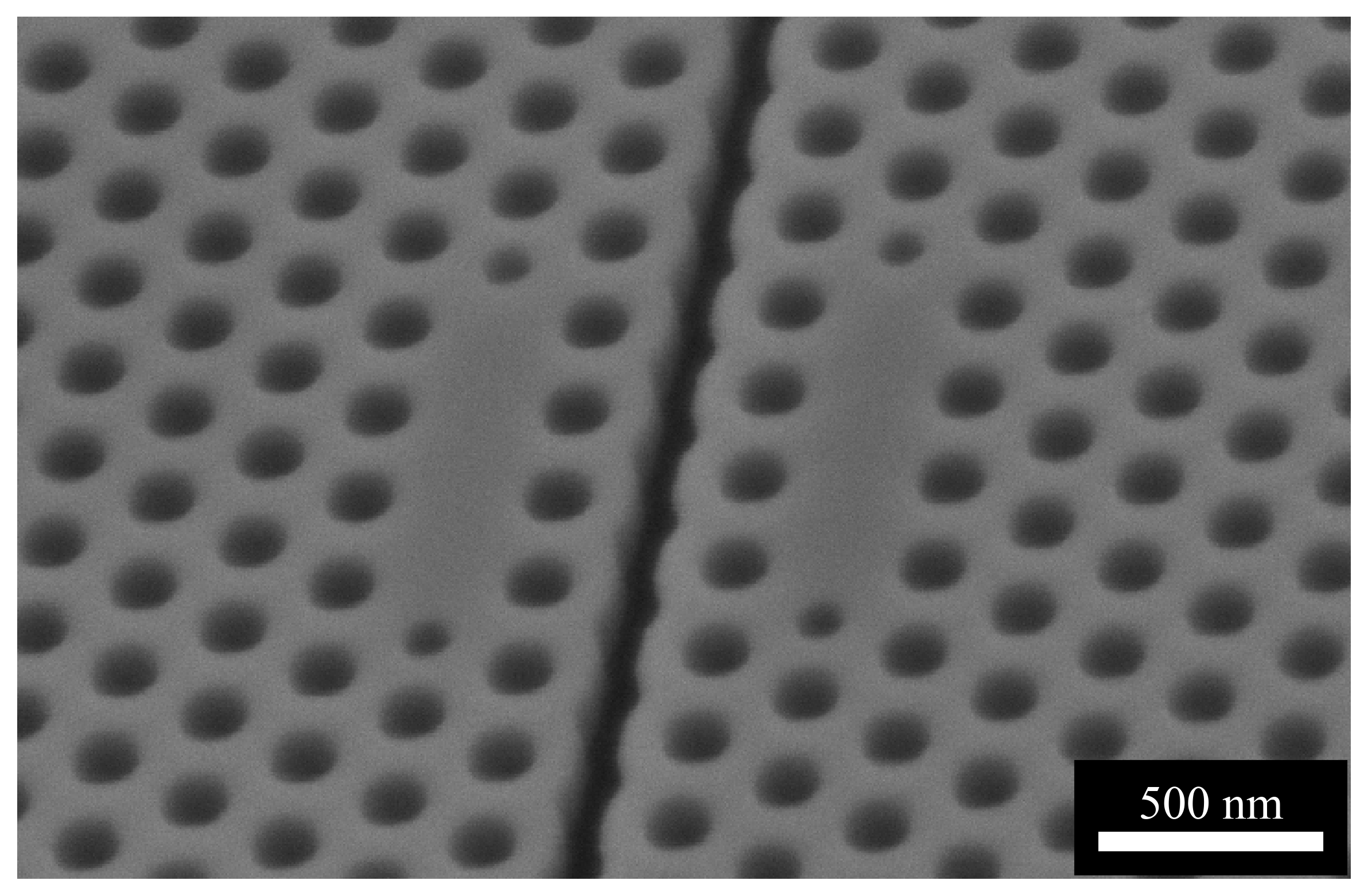} 
        \caption{}
        \label{fig:sem_1}
    \end{subfigure}
    \hfill
    \begin{subfigure}[htbp]{0.495\linewidth}
        \centering
        \includegraphics[width=\linewidth]{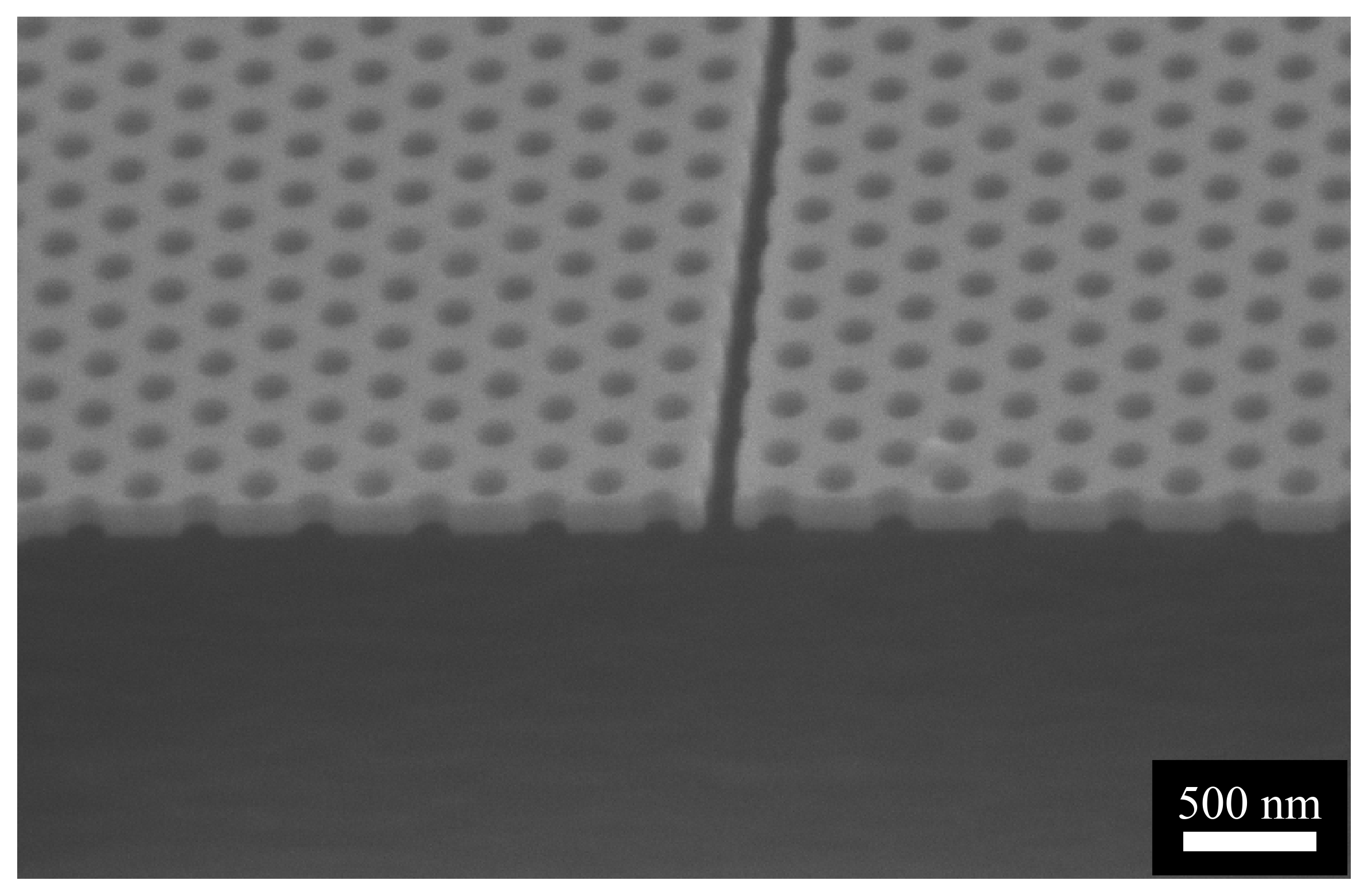}
        \caption{}
         \label{fig:sem_2}
    \end{subfigure}
    \caption{Tilted scanning electron micrograph showing (a) the etched electrical isolation trench and optimized holes adjacent to the cavity, (b) sidewalls of air holes and undercut of the split cavity. Scale bars are 500 nm.}
    \label{fig:sem}
\end{figure}

A detailed description of the sample growth and fabrication can be found in the Supporting Information. In brief, the patterns are written and transferred using electron-beam lithography and chlorine-based inductively coupled plasma (Cl-ICP) etching. The membrane is subsequently released using hydrofluoric acid. A digital cleaning process involving hydrogen peroxide and potassium hydroxide \cite{Midolo_soft-mask_2015, Carfagno_sleeve_2023} is then used to remove the resist and wet-etch residues. Finally, the sample is dried via carbon dioxide critical point drying. To ensure that we obtain a device in which the cavity resonance peak lies within the tuning range of our laser, we fabricate a series of devices with different lattice constant scaling and different hole radius bias. See the Supporting Information for details. Fig.~\ref{fig:sem} shows Scanning Electron Microscope (SEM) images of a representative fabricated device.

We utilize the cross-polarized resonant scattering technique to characterize the split cavities \cite{McCutcheon_resonant_2005, Rivoire_gallium_2008, Portalupi_planar_2010, Portalupi_delibrate_2011, Deotare_coupled_2009, Deotare_high_2009, Galli_light_2009, Liapis_onchip_2016}. We choose this technique because it allows us to measure the intrinsic cavity Q without fabricating waveguides coupled to the cavity, which would introduce an additional loss pathway. Because cross-polarized resonant scattering is primarily sensitive to the even mode \cite{Deotare_coupled_2009}, we fabricated and characterized the device with the  optimal $\Delta x$ and $\Delta r$ for maximum even mode Q. 

\begin{figure}[H]
    \centering
    \includegraphics[width=\textwidth]{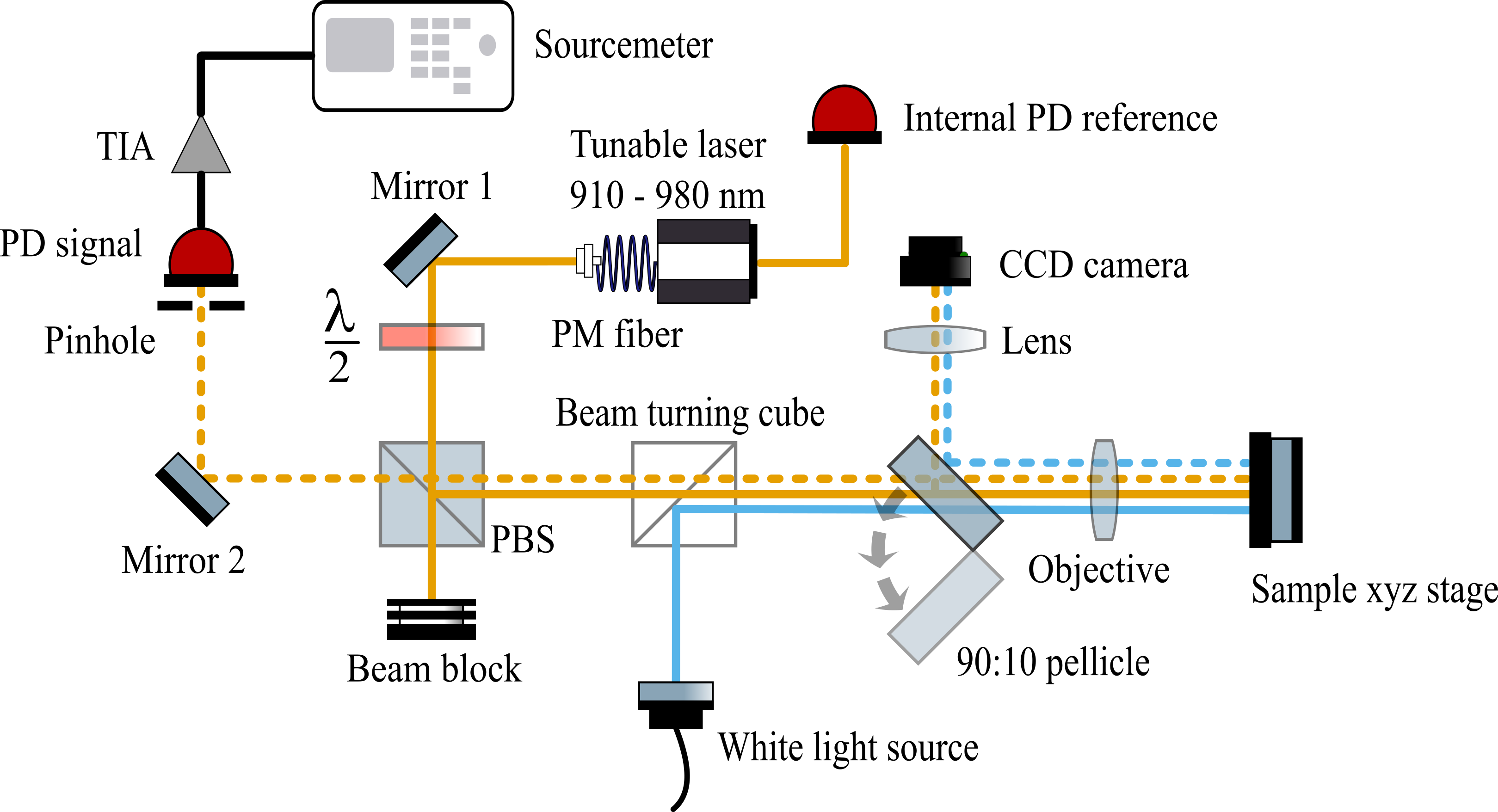}
    \caption{Schematic of the cross-polarized resonant scattering setup. PM: polarization-maintaining. $\lambda/2$: half-wave plate. PBS: polarizing beam splitter. CCD: charge-coupled device. PD: photodiode. TIA: trans-impedance amplifier. BTC: Beam Turning Cube used for white light illumination when imaging. Orange solid line: excitation laser beam. Orange dashed line: reflected laser beam from the sample. Blue solid line: white light illumination. Blue dashed line: reflected white light from the sample.}
    \label{fig:CRPS_diagram}
\end{figure}

Light from a tunable laser (Toptica CTL DLC 950, 910-980 nm tuning range) enters the setup via a polarization-maintaining fiber oriented for s-polarization in the experiment. A half-wave plate is used to fine-tune the alignment of the beam's polarization by maximizing the power of the s-polarized light reflected from the polarizing beam splitter. The s-polarized excitation beam is focused onto the center of the cavity using an NA = 0.65, 50x objective (Mitutoyo M Plan Apo NIR HR 50x). The sample is rotated so that the cavity mode is oriented at $45^\circ$ with respect to the s-polarized excitation beam. The p-polarized light scattered by the cavity resonance is collected by the same objective, passes through a polarizing beam splitter, and is detected by a silicon photodiode (Thorlabs DET36A2). This configuration maximizes the resonant scattering signal \cite{Portalupi_delibrate_2011}. The wavelength is scanned with a 10-pm step size and a 10 pm/s scan speed. Finally, the cavity signal is normalized by the laser's internal photodiode signal to eliminate laser power fluctuations. 

\section{Results and discussion}
We found that the fabricated split cavity whose lattice constant was scaled by 7\% and hole radius biased by -23 nm (SC23107) had a resonant peak positioned in the middle of the tuning range for our laser. We characterized 13 instances of this device. To extract the cavity Q, each scattering spectrum is fitted to a Fano line shape plus a linear offset to account for background reflections. This functional form is given by  Equation \ref{equ:fano},
\begin{align}
    F(E) &= a\frac{\left(q \frac{\Gamma}{2} + E - E_0\right)^2}{\left(\frac{\Gamma}{2}\right)^2 + (E - E_0)^2} + bE + c
    \label{equ:fano}
\end{align}
where E is the photon energy, $E_0$	is the resonant photon energy, $q$ is the Fano asymmetric parameter, $a$ is the Fano line shape amplitude, $\Gamma$ is the resonant linewidth, and $b, c$ are the slope and intercept of the offset spectrum, respectively. The cavity quality factor $Q$ is extracted via the relation $Q=E_0 / \Gamma,$ with its uncertainty calculated using the standard errors of the fitting parameters $E_0$ and $\Gamma$ along with their correlation. 

In principle Eqn.~\ref{equ:fano} can be fitted to the experimental data points $y_i$ by minimizing the sum of squared residuals:
\begin{align*}
    \min_{E_0, q, \Gamma, a, b, c} \sum_{i=1} [y_i - F(E_i)]^2.
\end{align*}
However, due to the nonlinearity of Eqn.~\ref{equ:fano}, the cavity Q extracted via such a direct fit depends heavily on the initial guess for each fitting parameter. We therefore employ the method proposed by Meierott et al.~\cite{Meierott_asymmetry_2016} to determine the optimal initial parameters before conducting the least-squares fit. Specifically, by implementing the following change of variables: $\varepsilon = (E-E_0) / \Gamma, k=a(q^2-1), l=2qa, m=b\Gamma, n=a+bE_0+c,$ Eqn.~\ref{equ:fano} can be rewritten as
\begin{align}
    F(\varepsilon) &= k \frac{1}{1 + \varepsilon^2} + l \frac{\varepsilon}{1 + \varepsilon^2} + m \varepsilon + n,
    \label{equ:fano_improved}
\end{align}
which reduces the number of nonlinear fitting parameters from three ($q, \Gamma, E_0$) to two ($\Gamma, E_0$).

The global minimum is then calculated using
\begin{align}
    \min_{E_0, q, \Gamma, a, b, c} \sum_{i=1} [y_i - F(E_i)]^2 \equiv \min_{E_0, \Gamma} \left[ \min_{k, l, m, n} \sum_{i=1} [y_i - F(\varepsilon_i)]^2 \right].
    \label{equ:lsq2}
\end{align}
Because Equation \ref{equ:fano_improved} is linear with respect to the fitting parameters k,l,m,n, the global minimum for the inner optimization over k,l,m,n can always be rigorously determined independently of the initial parameter guesses. The global minimum of Equation \ref{equ:lsq2} is then found by iteratively calculating the global minimum of its inner optimization over a predefined grid of $(\Gamma, E_0)$ pairs. Empirically, $E_0$ exerts less influence on the overall optimization landscape than $\Gamma$ because the value of $E_0$ is predominantly constrained by the peak signal intensity.

\begin{figure}[H]
    \centering
    \includegraphics[width=\linewidth]{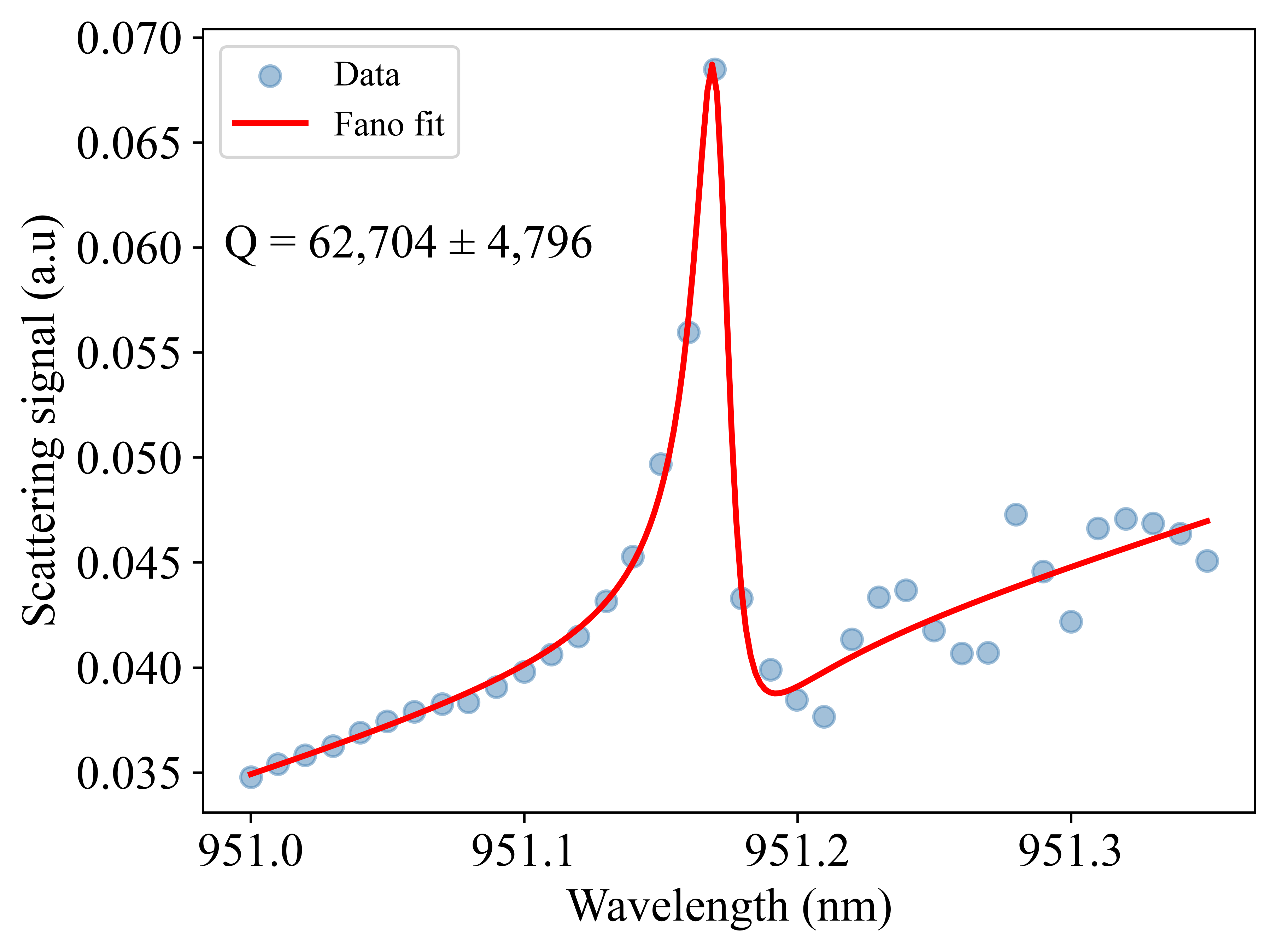}
    \caption{Resonant scattering spectrum and Fano fit for the split cavity device instance with the highest Q.}
    \label{fig:device20}
\end{figure}

\begin{figure}[H]
    \centering
    \includegraphics[width=\linewidth]{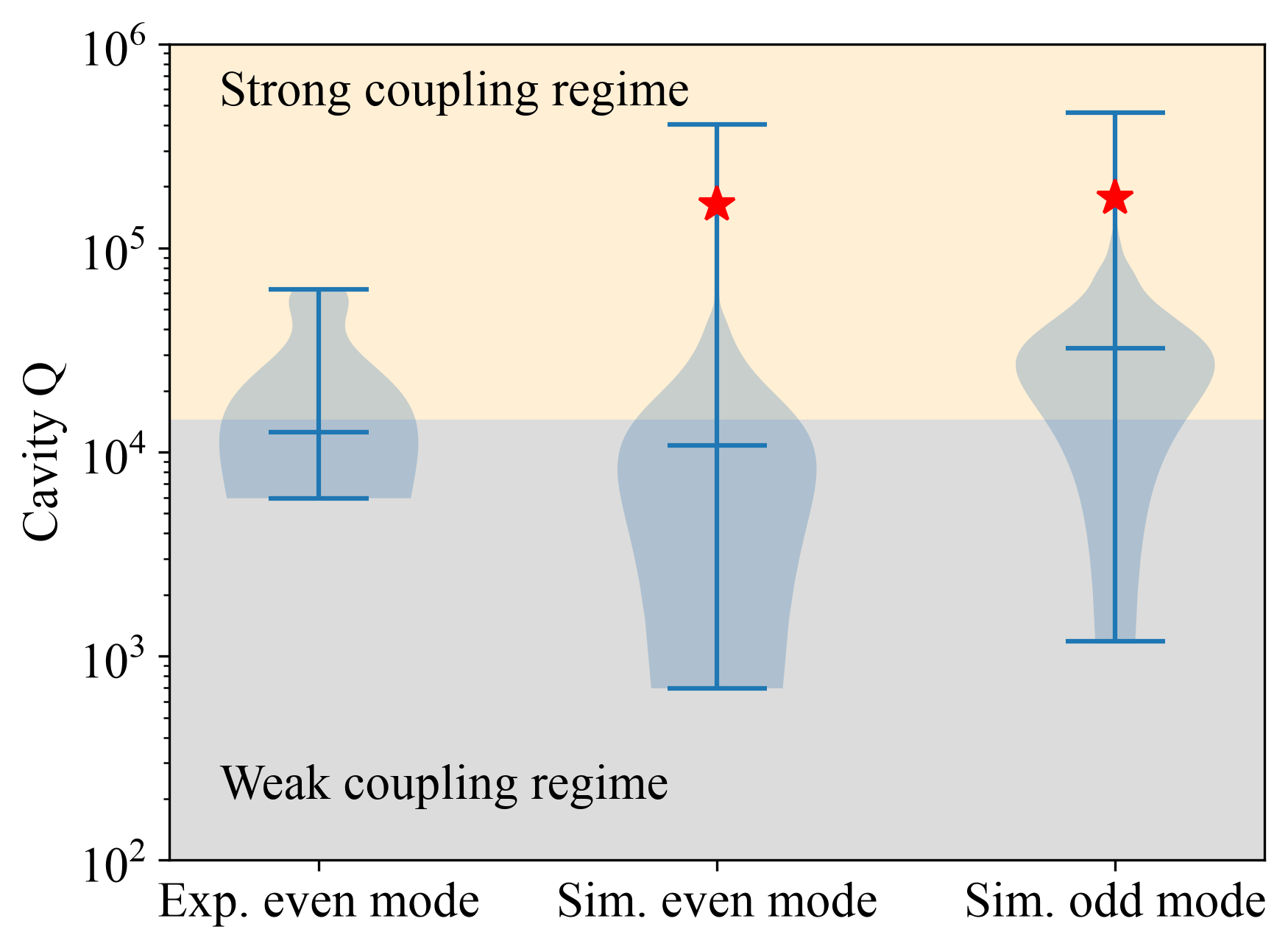}
    \caption{Violin plots comparing experimental cavity Qs of the even mode with Monte Carlo simulated cavity Qs of even and odd modes. The Monte Carlo sample size of each cavity mode is 530. The middle line in each violin plot marks its median, and the top and bottom lines mark its extrema. Red stars indicate the maximum computed Q for the even and odd modes obtained by sweeping the position and radius of the end holes, as reported in Fig \ref{fig:Q_delx_even_odd}.}
    \label{fig:violin_plot}
\end{figure}

Fig.~\ref{fig:device20} shows the resonant scattering spectrum and the corresponding fit for the split cavity that exhibited the highest quality factor of 62,704 $\pm$ 4,796. The data and fits for all other characterized devices can be found in the Supporting Information. Table \ref{tbl:summary_Q} summarizes the measured Q factor for all 13 device instances and the left column of Fig.~\ref{fig:violin_plot} reports this data as a violin plot. The mean and median cavity Q across the 13 device instances are $21,886 \pm 600$ and 12,417, respectively. 

To determine whether this Q is sufficient to enter the strong coupling regime, we look to prior work by Kim et al., who demonstrated strong coupling between \ce{InAs} quantum dots and an L3 cavity with a quality factor of 9,000 \cite{Kim_strong_2011}. The mode volume (V) of the split cavity is roughly double that of a standard L3 cavity. Because the cavity-quantum dot coupling strength, $g$, is inversely proportional to the square root of the mode volume $(g \propto 1/ \sqrt{V})$ \cite{Khitrova_vacuum_2006} and the cavity decay rate $\kappa$ is inversely proportional to the quality factor $(\kappa \propto 1/Q)$, we estimate that the split cavity requires a minimum quality factor of $9,000 \sqrt{2} \approx 12,728$ to reach the strong-coupling regime. This threshold is depicted by the shaded background in Fig.~\ref{fig:violin_plot}. The median Q for the experimentally-measured split cavity devices sits right on this threshold, with 6 of the measured devices having Qs exceeding the strong coupling threshold.

\begin{table}[H]
    \centering
    \begin{tabular}{|l|l|}
        \hline
        Device label & Cavity Q \\
        \hline
        SC23107, device 5  & 12,417 $\pm$ 1,142  \\
        \hline
        SC23107, device 6  & 7,187 $\pm$ 408  \\
        \hline
        SC23107, device 8  & 55,475 $\pm$ 3,818  \\
        \hline
        SC23107, device 9  & 25,152 $\pm$ 1,959  \\
        \hline
        SC23107, device 10  & 8,095 $\pm$ 376  \\
        \hline
        SC23107, device 12  & 6,929 $\pm$ 338  \\
        \hline
        SC23107, device 13  & 18,482 $\pm$ 844  \\
        \hline
        SC23107, device 14  & 5,940 $\pm$ 318  \\
        \hline
        SC23107, device 16  & 14,208 $\pm$ 1,077  \\
        \hline
        SC23107, device 17  & 10,917 $\pm$ 303  \\
        \hline
        SC23107, device 18  & 47,982 $\pm$ 3,214  \\
        \hline
        SC23107, device 20  & 62,704 $\pm$ 4,796  \\
        \hline
        SC23107, device 21  & 9,037 $\pm$ 317  \\
        \hline
    \end{tabular}
    \caption{Q factor and resonant wavelength of all 13 device instances}
    \label{tbl:summary_Q}
\end{table}

We next use simulations to understand how fabrication disorder contributes to the range of measured cavity Q values, which span 5,940 to 62,704. As described in the Supporting Information, we first obtain an experimental measure of our fabrication precision by analyzing a scanning electron micrograph of a reference two-dimensional hexagonal lattice photonic crystal that was fabricated simultaneously with the split cavities. We find that the fabricated holes have a mean radius of 73.4 nm with a standard deviation of 1.1 nm, a horizontal placement error with Gaussian width of 4.7 nm, and a vertical placement error with Gaussian width of 7.0 nm. We also find that the average fabricated cut width is 104.5 nm (14.5 nm larger than designed). We then conduct 530 FDTD simulations following the typical Monte Carlo approach: for each simulation the positional placement error and radial size error of each hole, relative to the design optimized for even-mode cavity Q, is drawn from the Gaussian distributions measured experimentally. The results of this analysis are shown in the middle column of Fig.~\ref{fig:violin_plot}. The distribution and median of the Q simulated with realistic fabrication disorder (10,796) is in good agreement with that measured experimentally (12,417). This similarity confirms that the experimentally measured cavity quality factor is primarily limited by positional and feature width disorders, which arise primarily from the precision of our e-beam lithography tool and uncertainty in feature bias due to wet etch. This implies that scattering and loss due to sidewall roughness or non-verticality, which are not included in our simulations, do not contribute significantly to the experimentally-measured Q values. This is consistent with Fig.~\ref{fig:sem_2}, which shows that our fabrication process results in hole sidewalls that are relatively vertical and smooth.

Finally, we perform another 530 Monte Carlo FDTD simulations to compute the expected Q of devices optimized for odd-mode Q. The results, shown in the right hand column of Fig.~\ref{fig:violin_plot}, demonstrate that a fabricated cavity optimized for odd mode Q can be expected to have a median cavity Q of 32,261. This is because, as illustrated in Fig.~\ref{fig:Ey_real}, the odd mode has less spatial overlap between its electric field and the disordered lattice than the even mode, particularly near the monolithic cut. Consequently, the fabrication-induced lattice disorder induces less loss for the odd mode. Importantly, the expected cavity Q for odd-mode-optimized devices has a distribution concentrated well within the strong-coupling regime even in the presence of realistic fabrication disorder. 

\section{Conclusion}
We presented the design, fabrication, characterization, and modeling of a split cavity embedded in a diode structure. We show that experimentally-fabricated devices optimized for even-mode cavity Q can achieve a median Q of 12,417, near the threshold for entering the strong coupling regime. Our analysis establishes that the Q is limited primarily by unavoidable fabrication disorder, rather than scattering. We further show that split cavity devices optimized for odd-mode Q can be expected to have a median Q lying comfortably within the strong coupling regime, even in the presence of realistic fabrication disorder. This split cavity device therefore presents a promising path toward devices in which strong coupling to the cavity mode can be used to mediate interactions between two quantum dots while simultaneously enabling independent Stark tuning of each individual quantum dot. Consequently, this split cavity design provides an important tool for the generation of photonic graph states for fully on-chip quantum information processing.

\nocite{*}

\section*{Acknowledgments}
The authors thank MS Parasnis for assistance with X-ray diffraction. The authors acknowledge support from the National Science Foundation (2217786). This work is supported by the University of Delaware Graduate College through L.H.M.s Unidel Distinguished Graduate Scholar Award. Any opinions, findings, and conclusions or recommendations expressed in this material are those of the authors(s). 


\section*{Supporting information}

\subsection*{Sample growth and material characterization}
The diode structure was grown on a semi-insulating, single-side-polished (100) GaAs substrate by molecular beam epitaxy (MBE) using a Veeco genXPLOR MBE system. The substrate oxide was thermally desorbed at $620^{\circ}$C under an As overpressure of $10^{-5}$ Torr beam-equivalent pressure (BEP). The \ce{Al_{0.75}Ga_{0.25}As} sacrificial layer and the p–i-p-i–n structure were grown in a single continuous run at a substrate temperature of $580^{\circ}$C. The substrate temperature during growth is measured using a kSA BandiT band edge thermometry (BET) system from k-Space Associates, with a broadband white light source and a detector mounted on separate viewports of the MBE growth chamber. Following growth of a 100 nm GaAs buffer layer, a 1370-nm \ce{Al_{0.75}Ga_{0.25}As} (x = $0.75 \pm 0.05$) sacrificial layer was deposited, followed by the diode structure, comprising in growth order: a 13.8-nm n-GaAs layer (silicon (Si)-doped, $8 \times 10^{18}$ cm$^{-3}$), a 13.8-nm n-GaAs layer (Si-doped, $2 \times 10^{18}$ cm$^{-3}$), an 18.9-nm intrinsic GaAs spacer, an 11.1-nm p-GaAs layer (beryllium (Be)-doped, $2 \times 10^{18}$ cm$^{-3}$), a 55.9-nm intrinsic GaAs spacer, a 22.6-nm p-GaAs layer (Be-doped, $2 \times 10^{18}$ cm$^{-3}$), and a 13.8-nm p-GaAs layer (Be-doped, $8 \times 10^{18}$ cm$^{-3}$). The total diode thickness is $149.9 \pm 2$ nm.

\begin{figure}[H]
    \centering
    \includegraphics[width=0.8\linewidth]{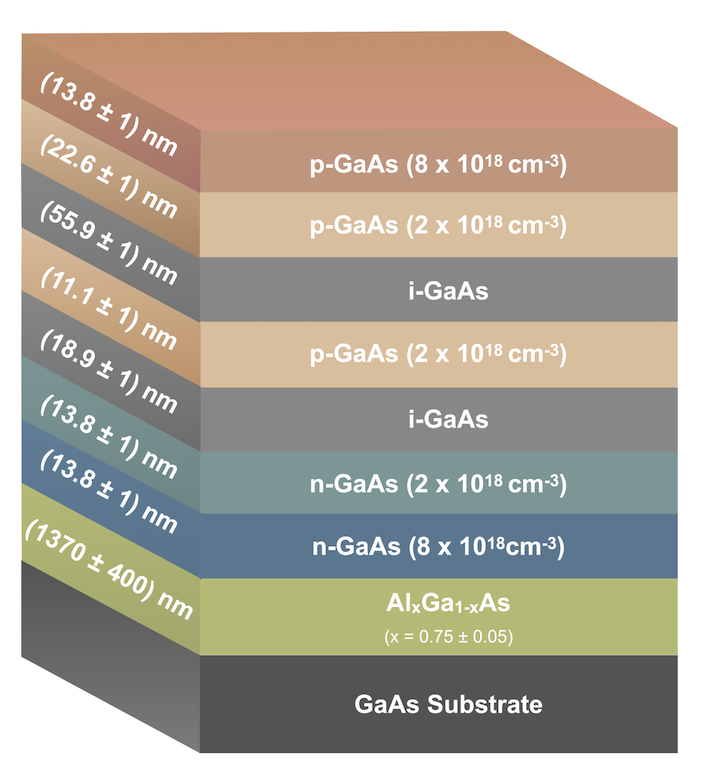}
    \caption{Material stack showing p-i-p-i-n diode embedded in the GaAs membrane. Not to scale.}
    \label{fig:pipin-stack}
\end{figure}

\begin{figure}[H]
    \centering
    \includegraphics[width=\textwidth]{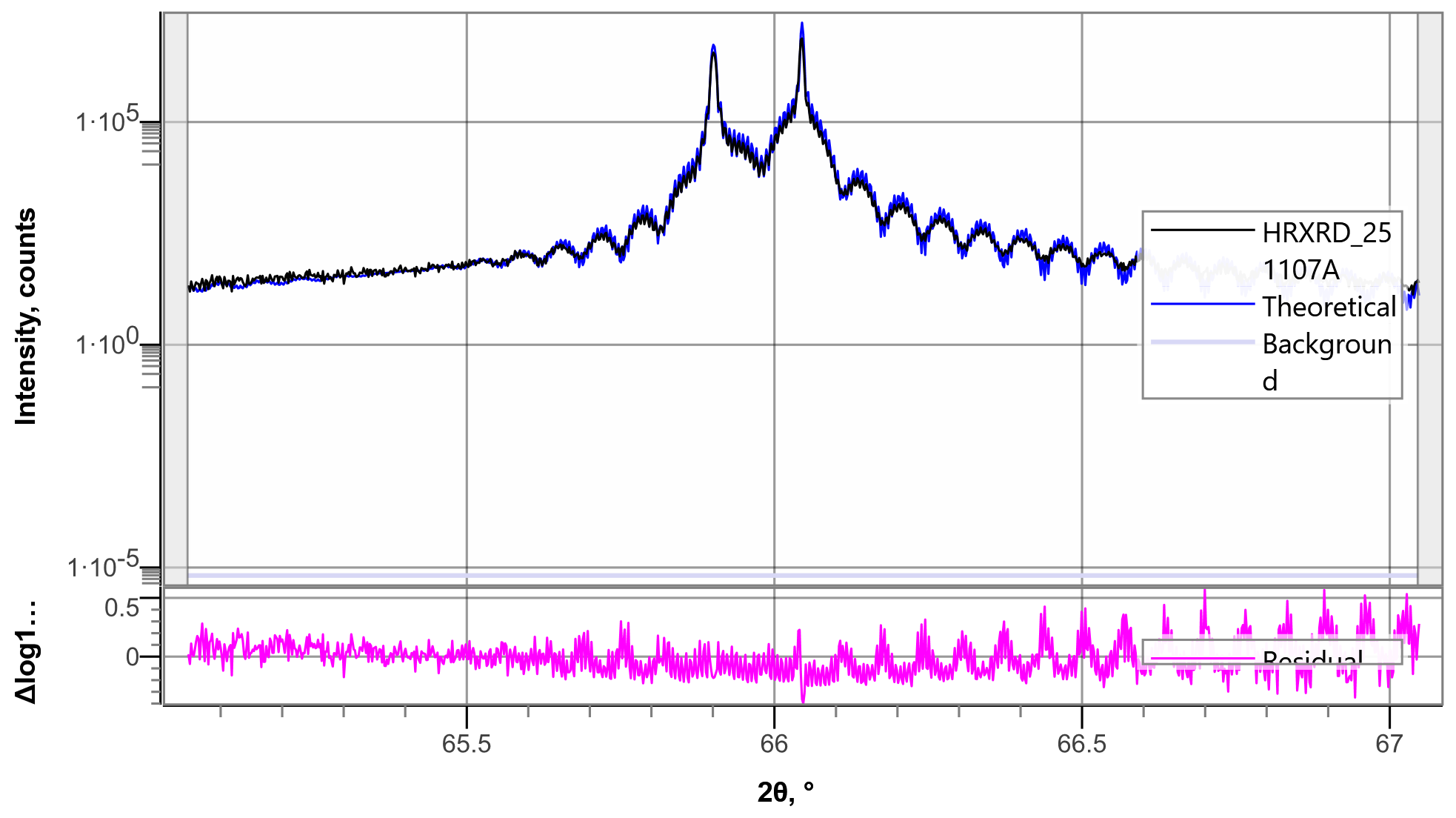}
    \caption{Rocking curve XRD result. The top panel shows the measurement data (blue) along with a theoretical fit (black). The bottom panel shows the difference between the data and theoretical fit (magenta). A constant background is assumed.}
    \label{fig:xrd}
\end{figure}

Prior to device fabrication, film thicknesses and the aluminum (\ce{Al}) concentration in the \ce{Al_{x}Ga_{1-x}As} are measured by rocking curve X-ray diffraction (XRD). Fig.~\ref{fig:xrd} shows the XRD result of the sample used for device measurements. The peak at $2\theta = 65.900^\circ$ is associated with \ce{Al_x Ga_{1-x}As} and the peak at $2\theta = 66.043^\circ$ is associated with \ce{GaAs}. The data is modeled with three fitting parameters: we find a \ce{GaAs} membrane thickness of $161.21 \pm 0.04$ nm, a \ce{Al_x Ga_{1-x}As} thickness of $1420.92 \pm 0.13$ nm, and \ce{Al} concentration concentration of $x$ is $75.69\pm 0.01\%.$

\subsection*{Fabrication process details}
We next proceed to fabricate the devices using the following recipe: 
\begin{enumerate}[noitemsep]
    \item Spin AR-P6200.09 e-beam resist at 3000 rpm for 1 minute. Soft bake at $170^\circ$C for 5 minutes.
    \item Electron beam lithography (Raith EBPG 5200): 100 kV accelerating voltage, 1 nA current, 300 $\mu$m aperture, and 200 $\mu$C/cm$^2$ dose. BEAMER parameters for GPF file processing are summarized in Table ~\ref{tbl:BEAMER}.
    \item Develop resist with AR 600-546 developer for 2 minutes. Rinse with isopropyl alcohol (IPA) and dry with nitrogen.
    \item Descum in oxygen plasma for 1 minute.
    \item RIE-ICP etch (PlasmaTherm Apex SLR): 15 sccm \ce{BCl3} and 10 sccm \ce{Ar} gas flow rates, 6 mTorr pressure, 500 W coil power, 25 W bias power, and 4 minutes etch time. The etch is divided into four 1-min etch cycles to minimize resist hardening from prolonged high temperature exposure. The resist is cooled for 1 minute between etch cycles.
    \item Strip resist: soak in N-Methylpyrrolidone at $80^\circ$C for 2 hours. Rinse with IPA and dry with nitrogen.
    \item Membrane release with hydroflouric acid (HF 5\%) wet etch for 2 minutes. The sample is kept under liquid after this etch to prevent capillary force damage from water evaporation.
    \item Digital cleaning to remove resist and HF etch residues: soak in hydrogen peroxide (\ce{H_2O_2} 30\%) for 1 minute, then rinse in de-ionized water for 1 minute, then soak in potassium hydroxide (\ce{KOH} 22.5\%) for 1 minute, then rinse in de-ionized water for 1 minute.
    \item Critical point drying (Tousimis Autosamdri-815, Series B) under the ``Delicate Sample" instructions.
\end{enumerate}

\begin{table}[H]
  \centering
  \begin{tabular}{|l|l|}
    \hline
    Parameter Name & Parameter Value \\
    \hline
    Writing Grid Resolution & 0.001 $\mu$m \\
    \hline
    Beam Step Size & 0.005 $\mu$m  \\
    \hline
    Mainfield Resolution & 0.0005 $\mu$m \\
    \hline
    Mainfield Size (X $\times$ Y) & 25 $\mu$m $\times$ 25 $\mu$m \\
    \hline
    Subfield Resolution & 0.0005 $\mu$m \\
    \hline
    Subfield Size (X $\times$ Y) & 2.5 $\mu$m $\times$ 2.5 $\mu$m \\
    \hline
    Fracturing Mode & Curved \\
    \hline
    Field Ordering & Manual \\
    \hline
    Feature Ordering & ArrayCompaction \\
    \hline
    Region Size & 25 $\mu$m \\
    \hline
    Region Traversal & SpiralOutwards \\
    \hline
    Exposure Dose & Ascending \\
    \hline
    Fracture Control & High Resolution Mode \\
    \hline
    Trapezoids & X and Y \\
    \hline
    Multipass & 1 \\
    \hline
  \end{tabular}
  \caption{BEAMER GenISys parameters}
  \label{tbl:BEAMER}
\end{table}

\subsection*{Quantifying fabrication disorder}
\begin{figure}[H]
    \centering
    \begin{subfigure}[htbp]{0.495\linewidth}
        \centering
        \includegraphics[width=\linewidth]{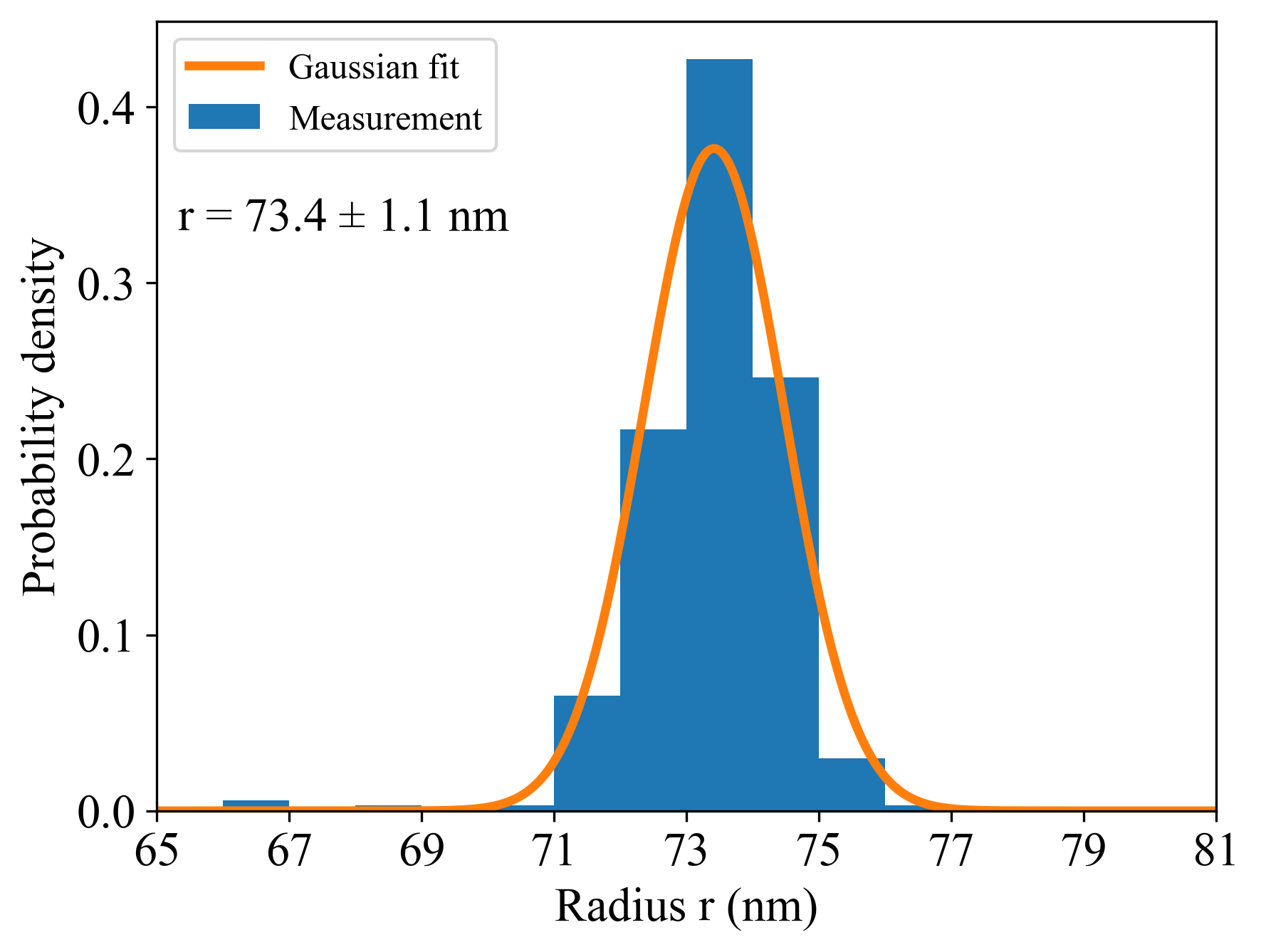}
        \caption{}
        \label{fig:r_dist}
     \end{subfigure}
    \hfill    
    \begin{subfigure}[htbp]{0.495\linewidth}
        \centering
        \includegraphics[width=\linewidth]{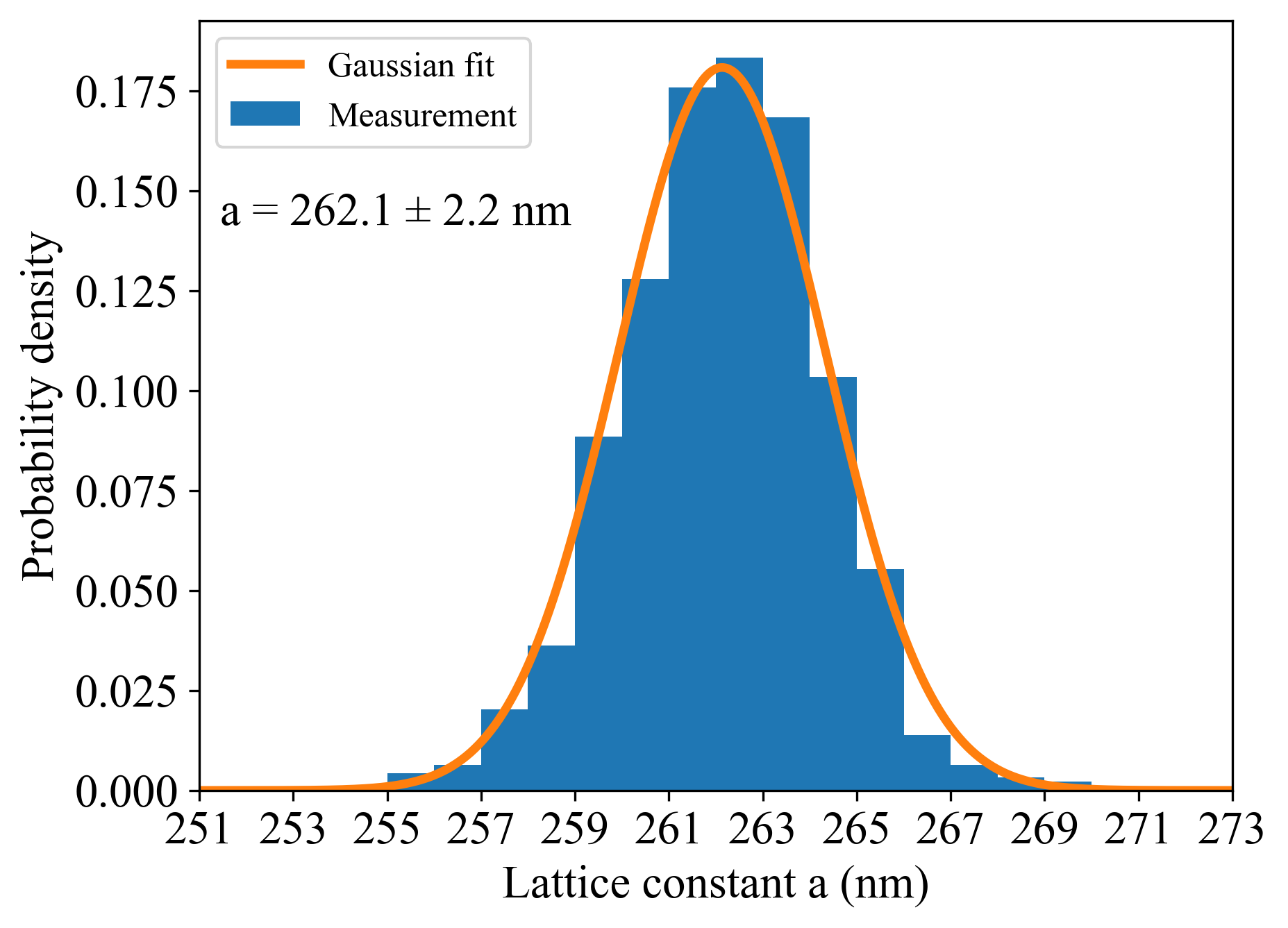}
        \caption{}
        \label{fig:a_dist}
    \end{subfigure}
    \caption{Histograms of radii (a) and lattice constants (b) extracted from scanning electron micrograph analysis. Each histogram is fitted to a Gaussian distribution. Fabrication errors are the standard deviations.}
    \label{fig:a_r_distrs}
\end{figure}

\begin{figure}[H]
    \centering
    \begin{subfigure}[htbp]{0.495\linewidth}
        \centering
        \includegraphics[width=\linewidth]{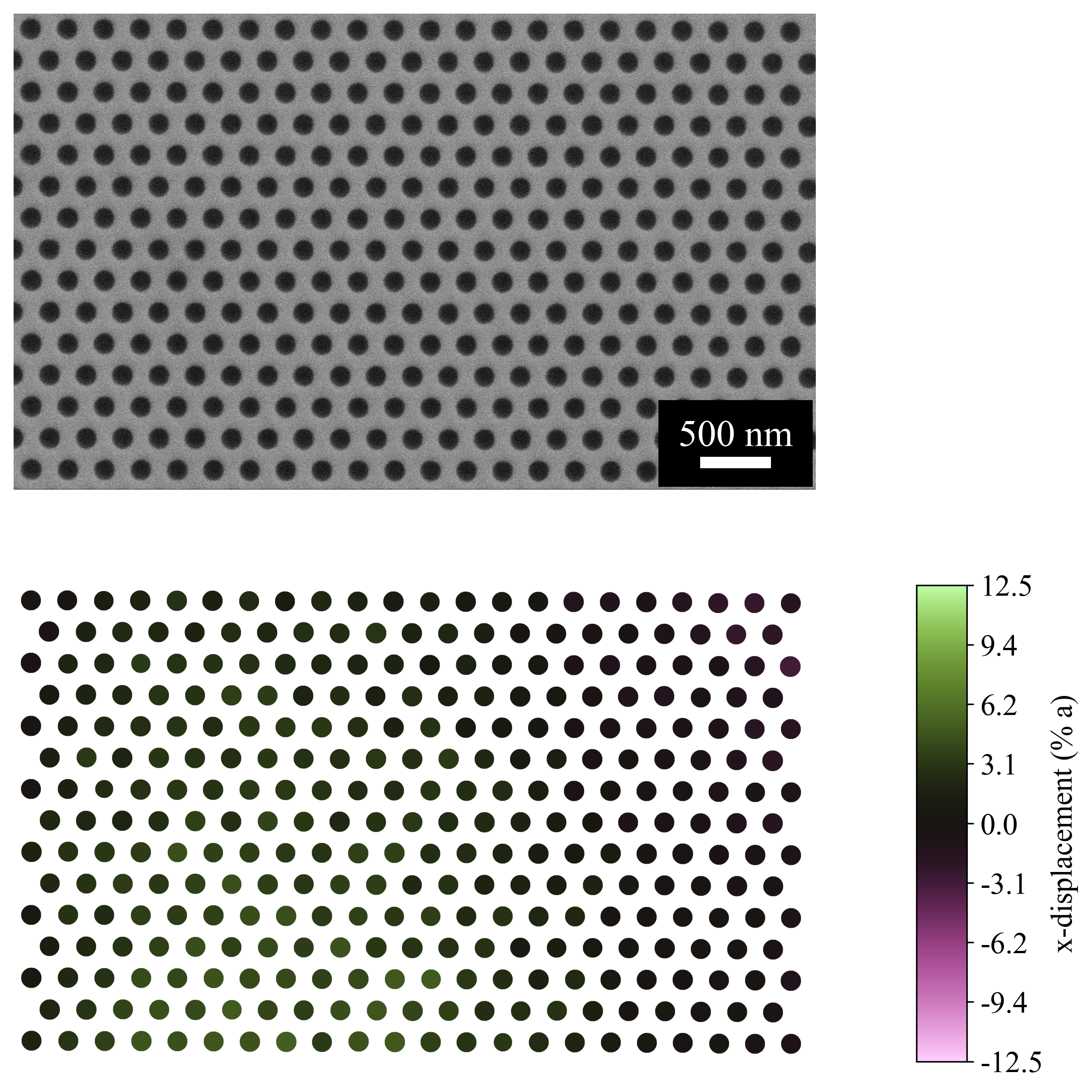} 
        \caption{}
    \end{subfigure}
    \hfill
    \begin{subfigure}[htbp]{0.495\linewidth}
        \centering
        \includegraphics[width=\linewidth]{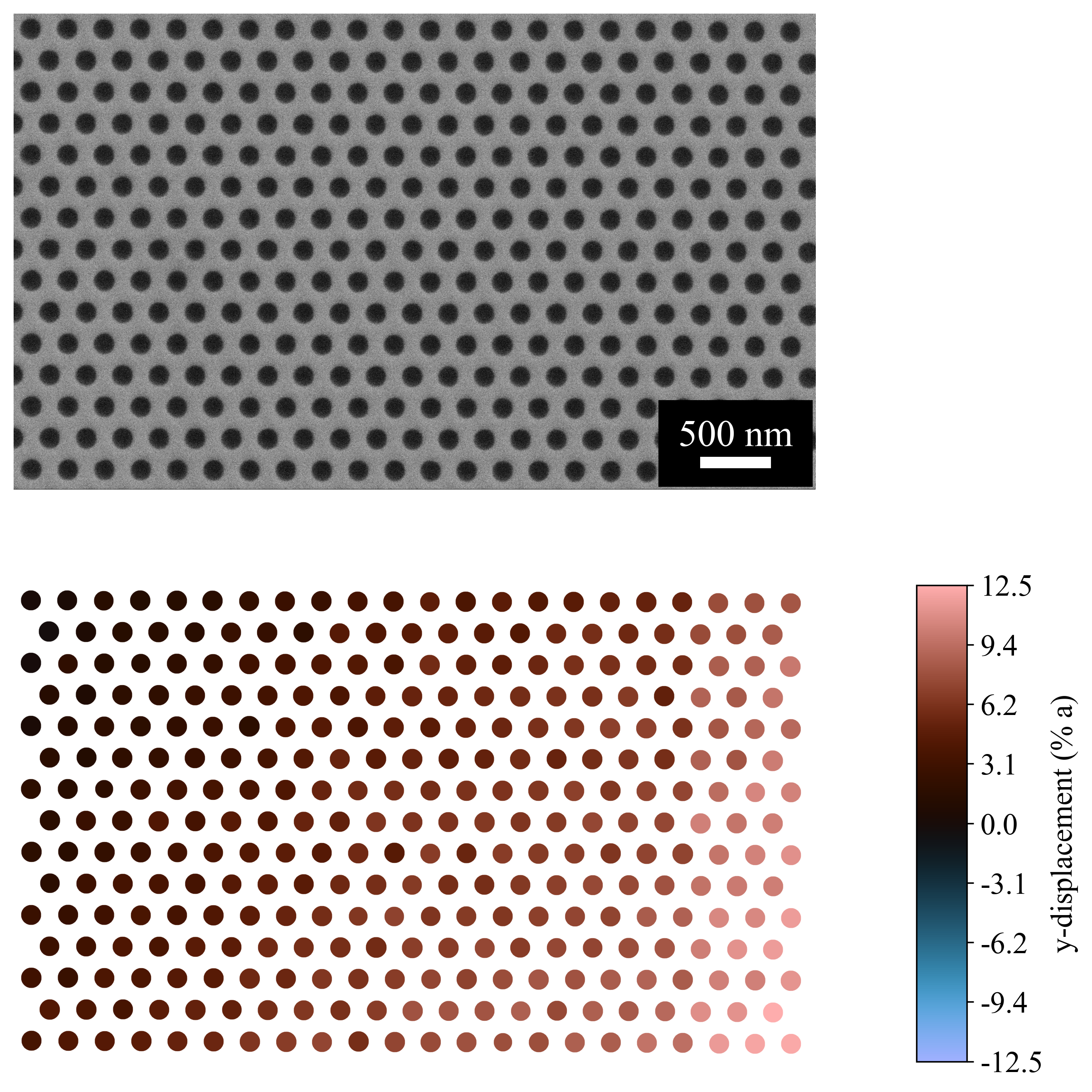}
        \caption{}
    \end{subfigure}
    \caption{Horizontal (a) and vertical (b) deviations of air holes from the nominal fabricated lattice constant of $a = 262.1$ nm.}
    \label{fig:disorder}
\end{figure}

To quantify fabrication disorder we first analyze a scanning electron micrograph of a reference two-dimensional hexagonal lattice photonic crystal that was fabricated simultaneously with the split cavities. As shown in Fig.~\ref{fig:r_dist}, the distribution of the measured radius of the air holes sampled from a subarea of the reference photonic crystal can be fit with a Gaussian to find an average hole radius of 73.4 nm and a standard deviation of 1.1 nm. We take this standard deviation as the measure of the radial fabrication disorder. Similarly, and as shown in Fig.~\ref{fig:a_dist}, the nearest-neighbor distance histogram of all air holes, which is again extracted from the SEM images, is fit to a Gaussian distribution to estimate the nominal lattice constant and standard deviation of $a = 262.1 \pm 2.2$ nm. The fitted nominal lattice constant is then used as the ideal baseline, allowing us to separately extract the horizontal $(\Delta x)$ and vertical $(\Delta y)$ deviation for each fabricated air hole, as shown in Fig.~\ref{fig:disorder}. We find a horizontal $(\Delta x)$ disorder of 4.7 nm and a vertical $(\Delta y)$ disorder of 7.0 nm.

\subsection*{Device measurements}
We cannot precisely predict the amount of isotropic etch bias (hole enlargement) that will occur during the \ce{Cl}-ICP and wet etch steps of a given fabrication run. Consequently, we fabricate a series of devices with the hole radius (excluding the optimized end-hole radius) biased, relative to the target size, by values ranging from -15 nm to -23 nm in -2 nm increments. We also cannot precisely know the index of refraction of the as-grown GaAs. Consequently, for each hole radius bias we fabricate devices with the lattice constant ($a$) scaled by 0\% to 10\% from the designed lattice constant ($a = 249$ nm) in 1\% increments. To determine which devices to use for characterization purposes, we focus first on identifying the correct hole radius bias. To do so, we analyze the SEM images of the reference photonic crystals (PhCs) fabricated simultaneously with the split cavities on the same sample. Each reference PhC has the same hole radius bias as a series of split cavity devices. As shown in Fig.~\ref{fig:r_dist}, hole radius bias of -23 nm results in the average radius of 73.4 nm, closest to the design value of 73 nm. Next, to determine the lattice constant that results in cavity resonance wavelength near the middle of our laser tuning range, we choose the series of devices with -23 nm hole radius bias and measure the cavity Q at increasing lattice constants. We find that the 7\% scale factor results in resonance wavelengths between 940 nm and 960 nm. Thus, we choose devices with -23 nm hole radius bias and lattice constant scaled by 7\% (SC23107) for our cavity Q measurements. The measurement of the device with the highest Q is reported in the main manuscript. Fig.~\ref{fig:devices_summary} shows the cavity Q measurements of the other 12 SC23107 device instances. The range of Q values obtained for these nominally identical devices is fully explained by the fabrication disorder, as explained in the main text. 

\begin{figure}[H]
    \centering
    \begin{subfigure}[htbp]{0.495\linewidth}
        \centering
        \includegraphics[width=\linewidth]{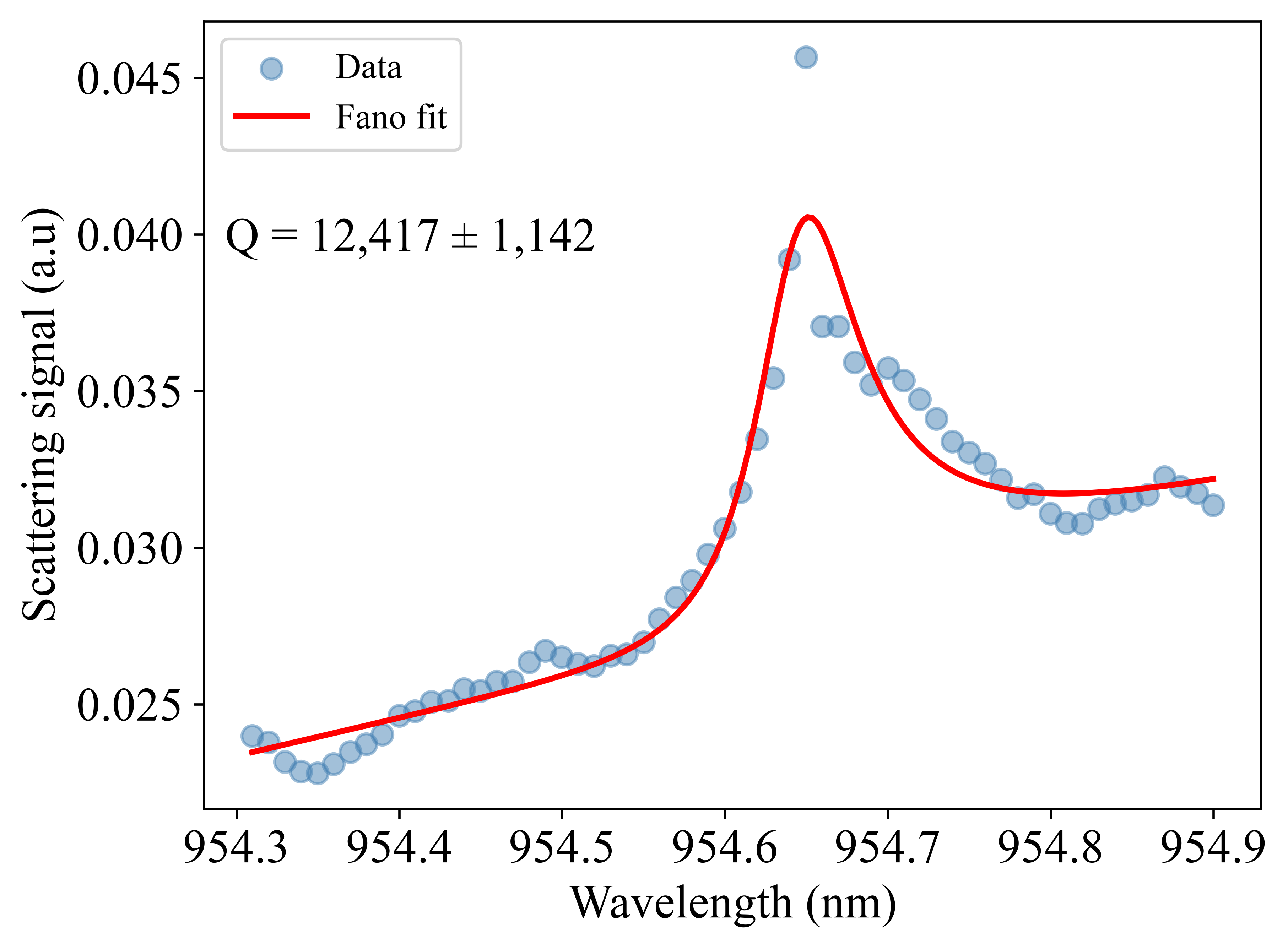} 
        \caption{SC23107, device 5}
        \label{fig:device5}
    \end{subfigure}
    \hfill
    \begin{subfigure}[htbp]{0.495\linewidth}
        \centering
        \includegraphics[width=\linewidth]{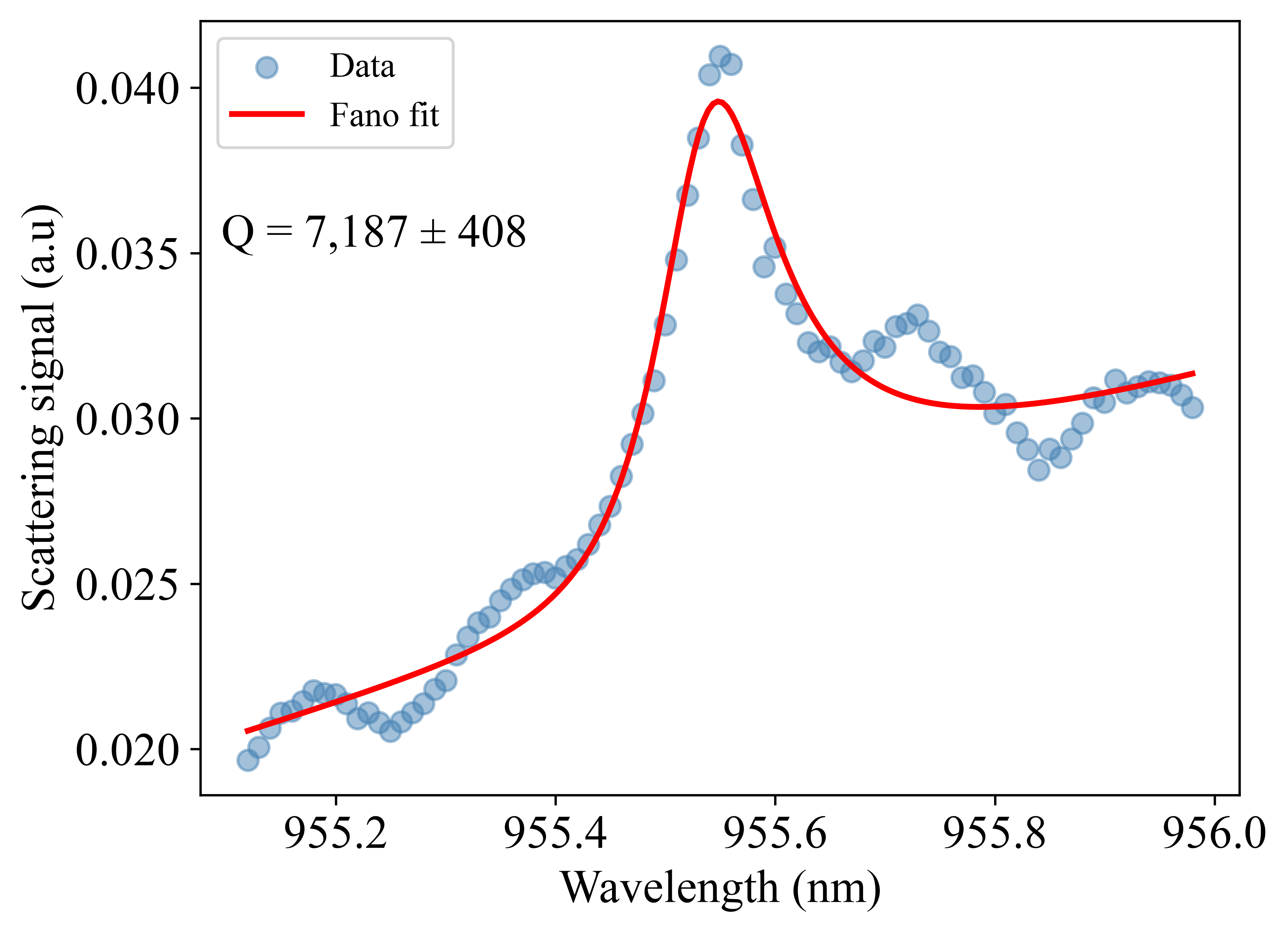}
        \caption{SC23107, device 6}
        \label{fig:device6}
    \end{subfigure}
    \\
    \begin{subfigure}[htbp]{0.495\linewidth}
        \centering
        \includegraphics[width=\linewidth]{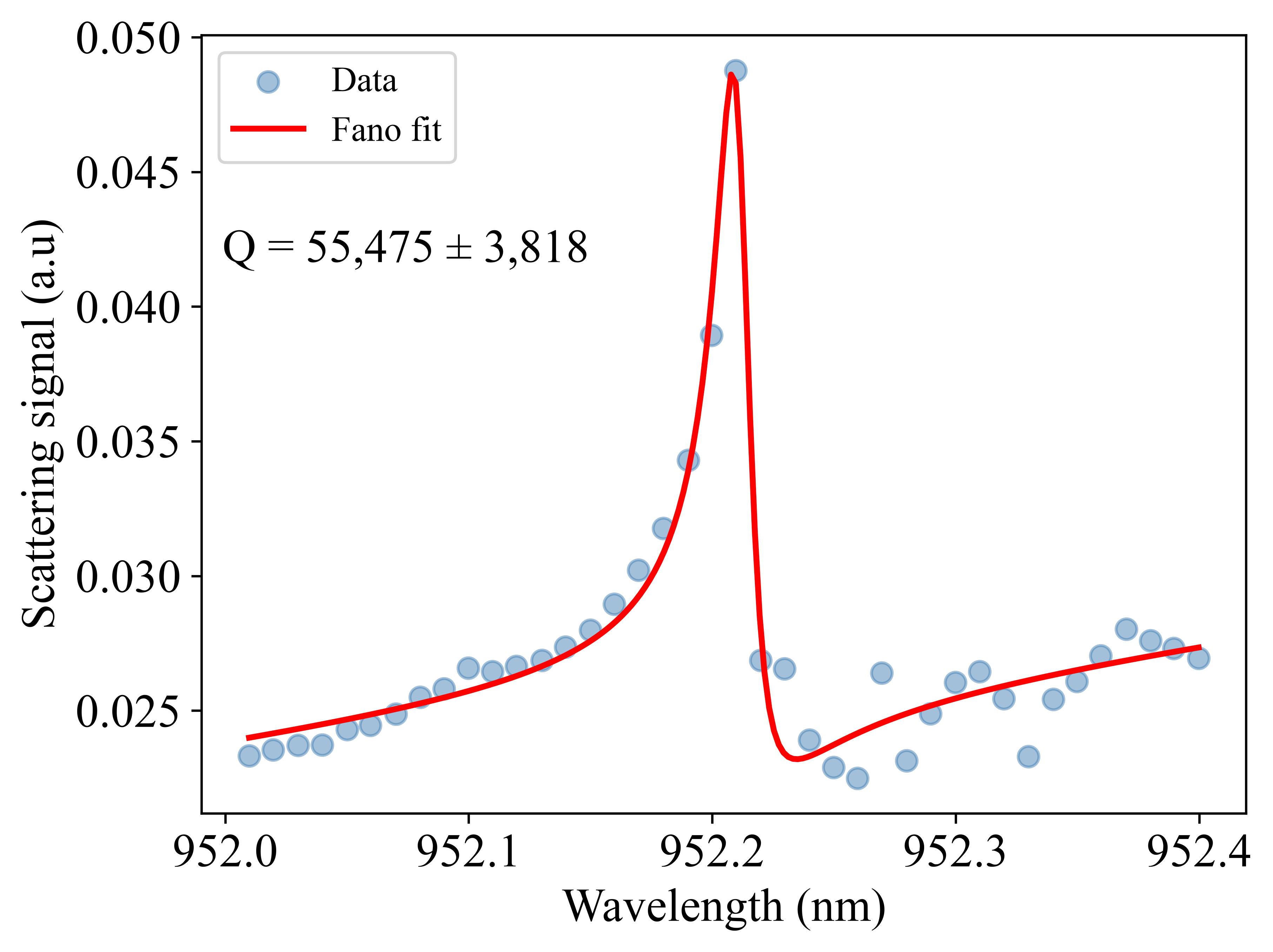} 
        \caption{SC23107, device 8}
        \label{fig:device8}
    \end{subfigure}
    \hfill
    \begin{subfigure}[htbp]{0.495\linewidth}
        \centering
        \includegraphics[width=\linewidth]{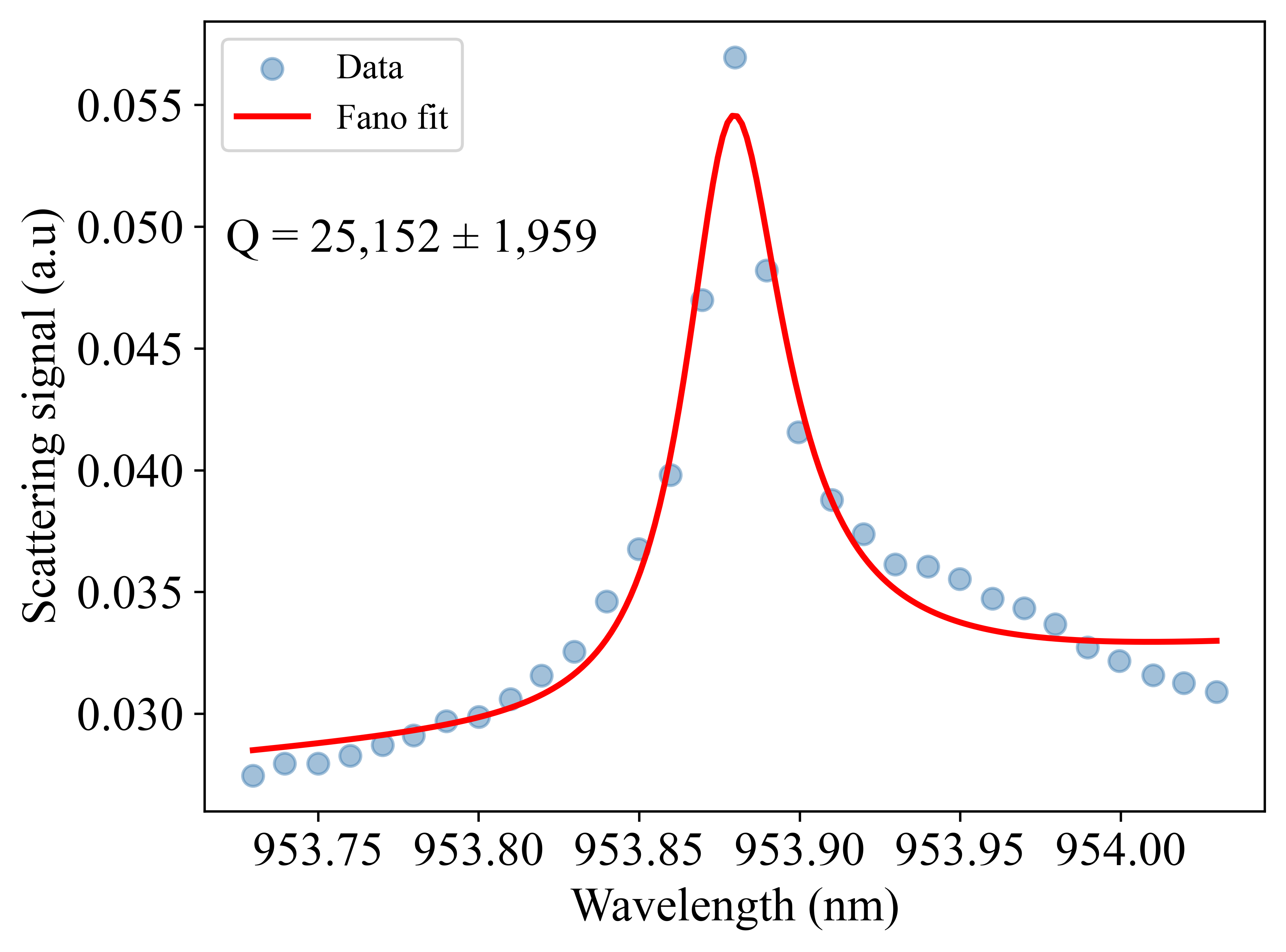}
        \caption{SC23107, device 9}
        \label{fig:device9}
    \end{subfigure}
    \\
    \begin{subfigure}[htbp]{0.495\linewidth}
        \centering
        \includegraphics[width=\linewidth]{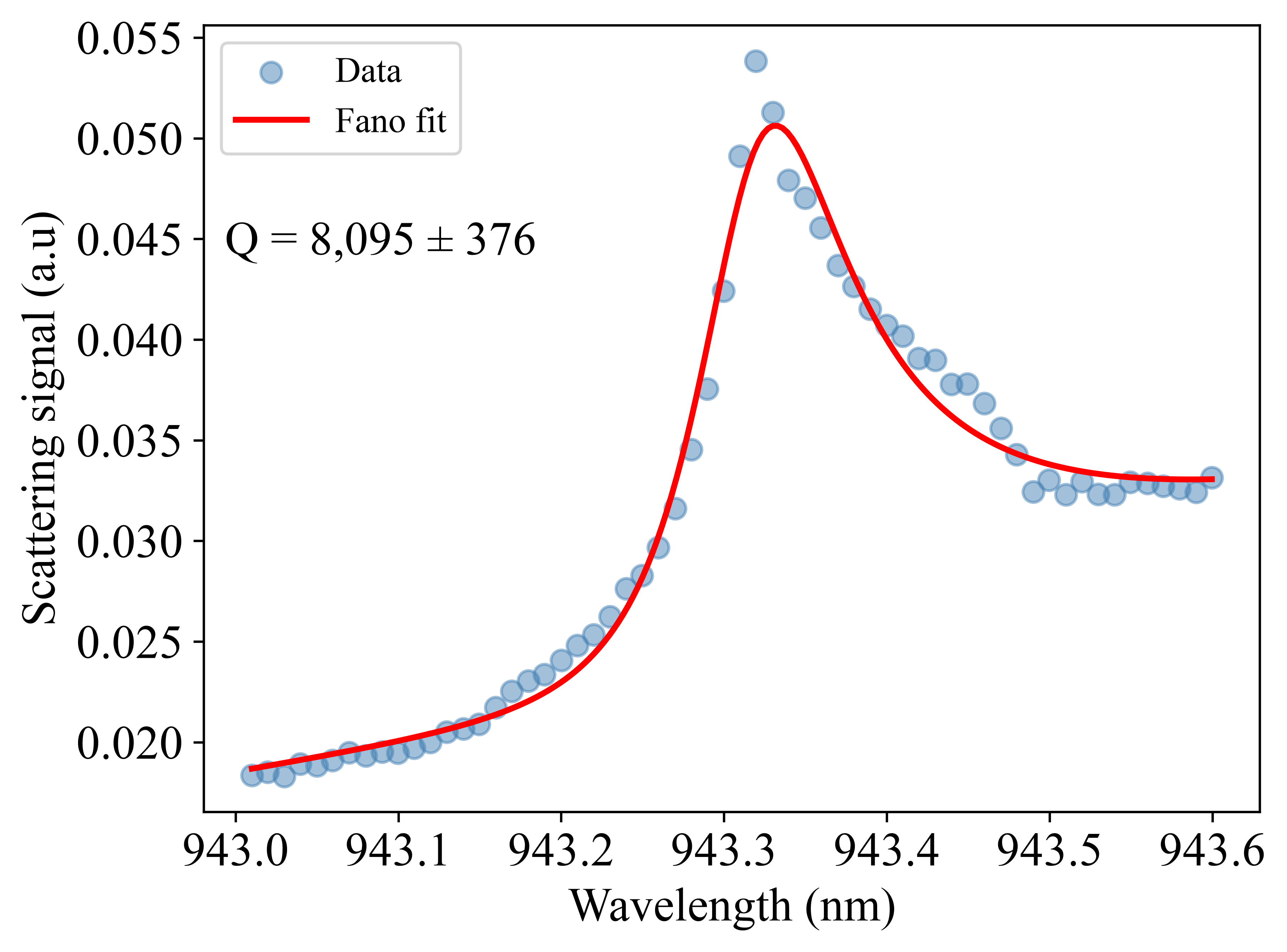} 
        \caption{SC23107, device 10}
        \label{fig:device10}
    \end{subfigure}
    \hfill
    \begin{subfigure}[htbp]{0.495\linewidth}
        \centering
        \includegraphics[width=\linewidth]{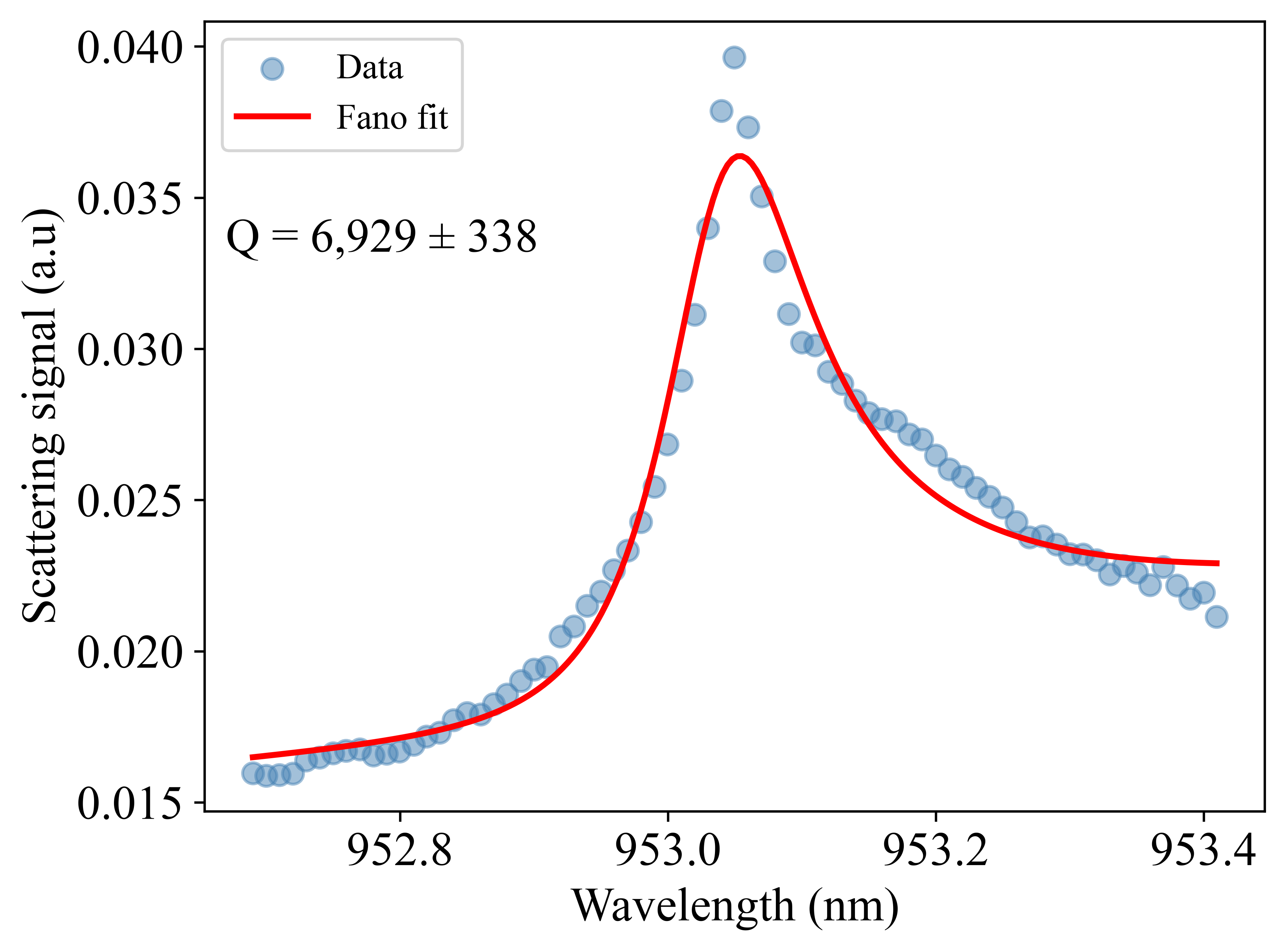}
        \caption{SC23107, device 12}
        \label{fig:device12}
    \end{subfigure}
\end{figure}

\begin{figure}[H]
    \ContinuedFloat
    \centering
    \begin{subfigure}[htbp]{0.495\linewidth}
        \centering
        \includegraphics[width=\linewidth]{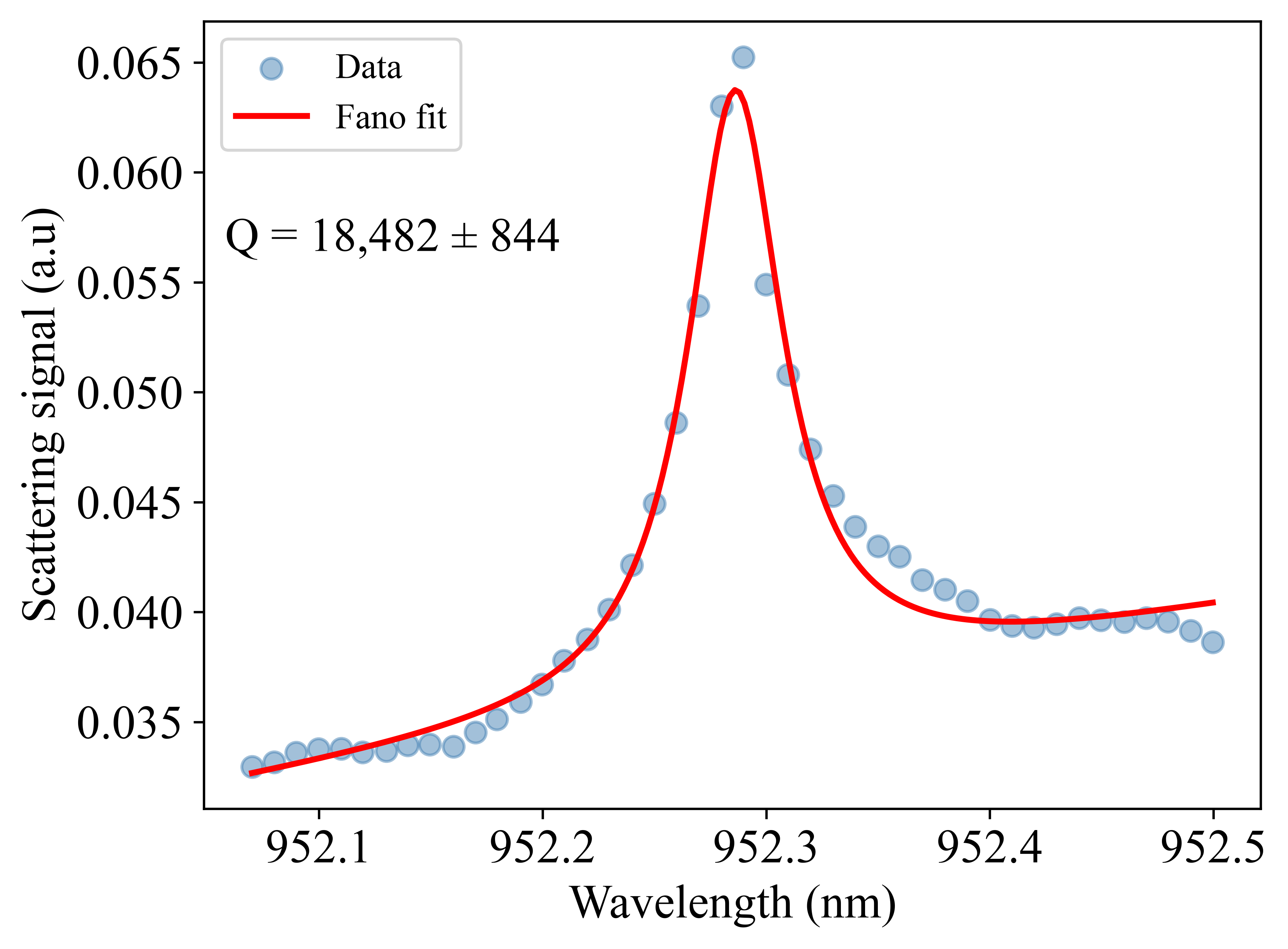} 
        \caption{SC23107, device 13}
        \label{fig:device13}
    \end{subfigure}
    \hfill
    \begin{subfigure}[htbp]{0.495\linewidth}
        \centering
        \includegraphics[width=\linewidth]{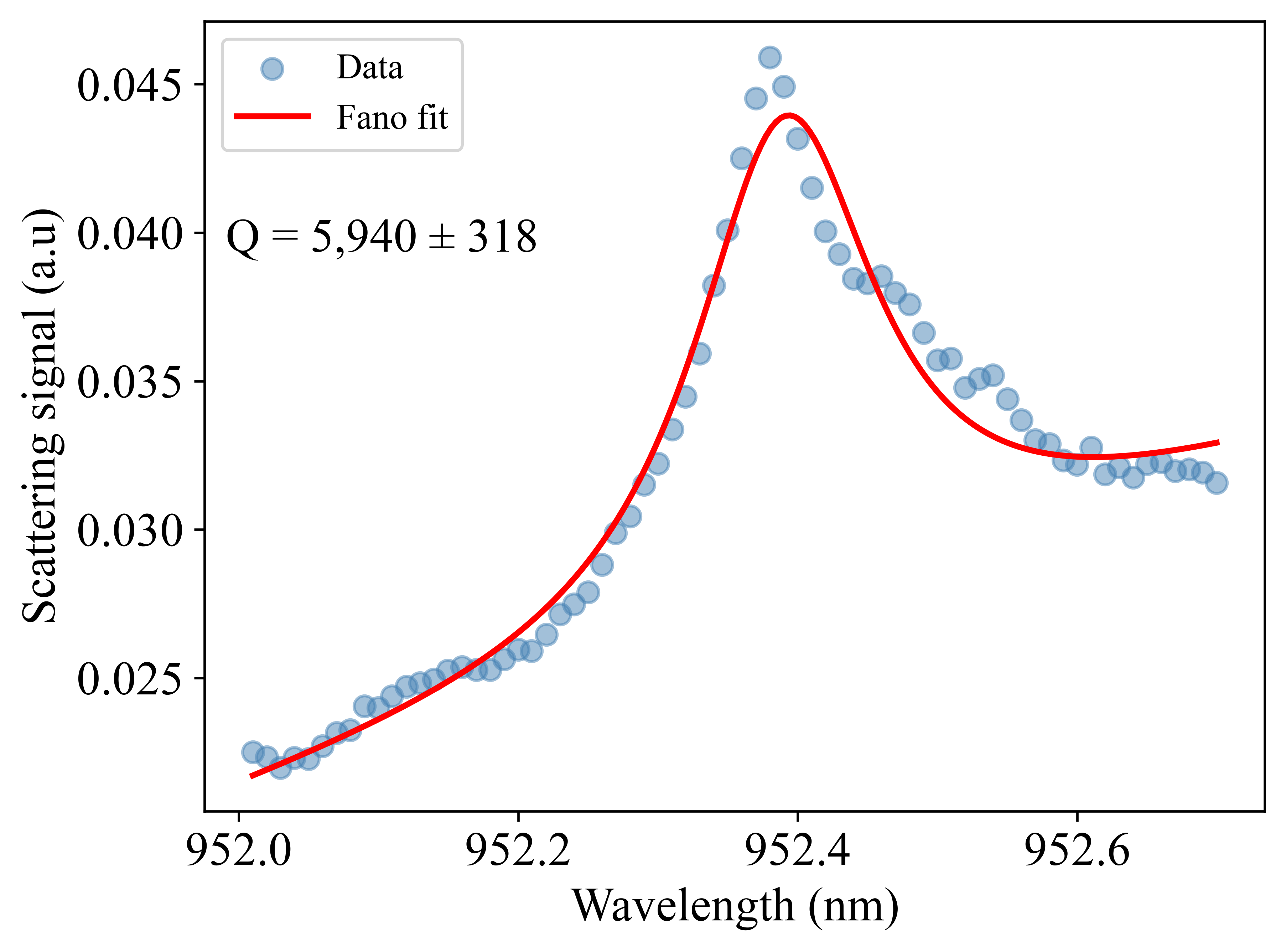}
        \caption{SC23107, device 14}
        \label{fig:device14}
    \end{subfigure}
    \\
    \begin{subfigure}[htbp]{0.495\linewidth}
        \centering
        \includegraphics[width=\linewidth]{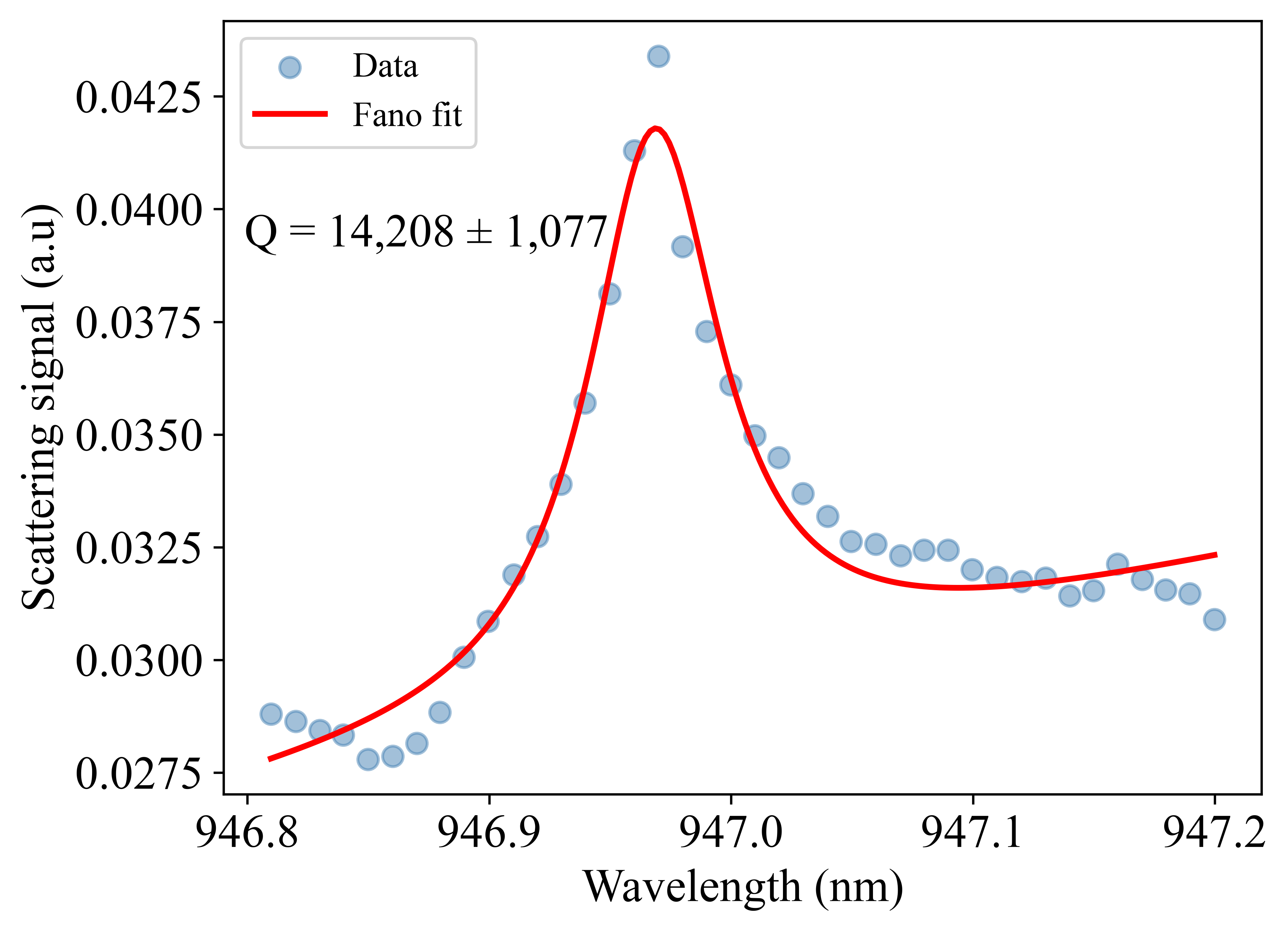} 
        \caption{SC23107, device 16}
        \label{fig:device16}
    \end{subfigure}
    \hfill
    \begin{subfigure}[htbp]{0.495\linewidth}
        \centering
        \includegraphics[width=\linewidth]{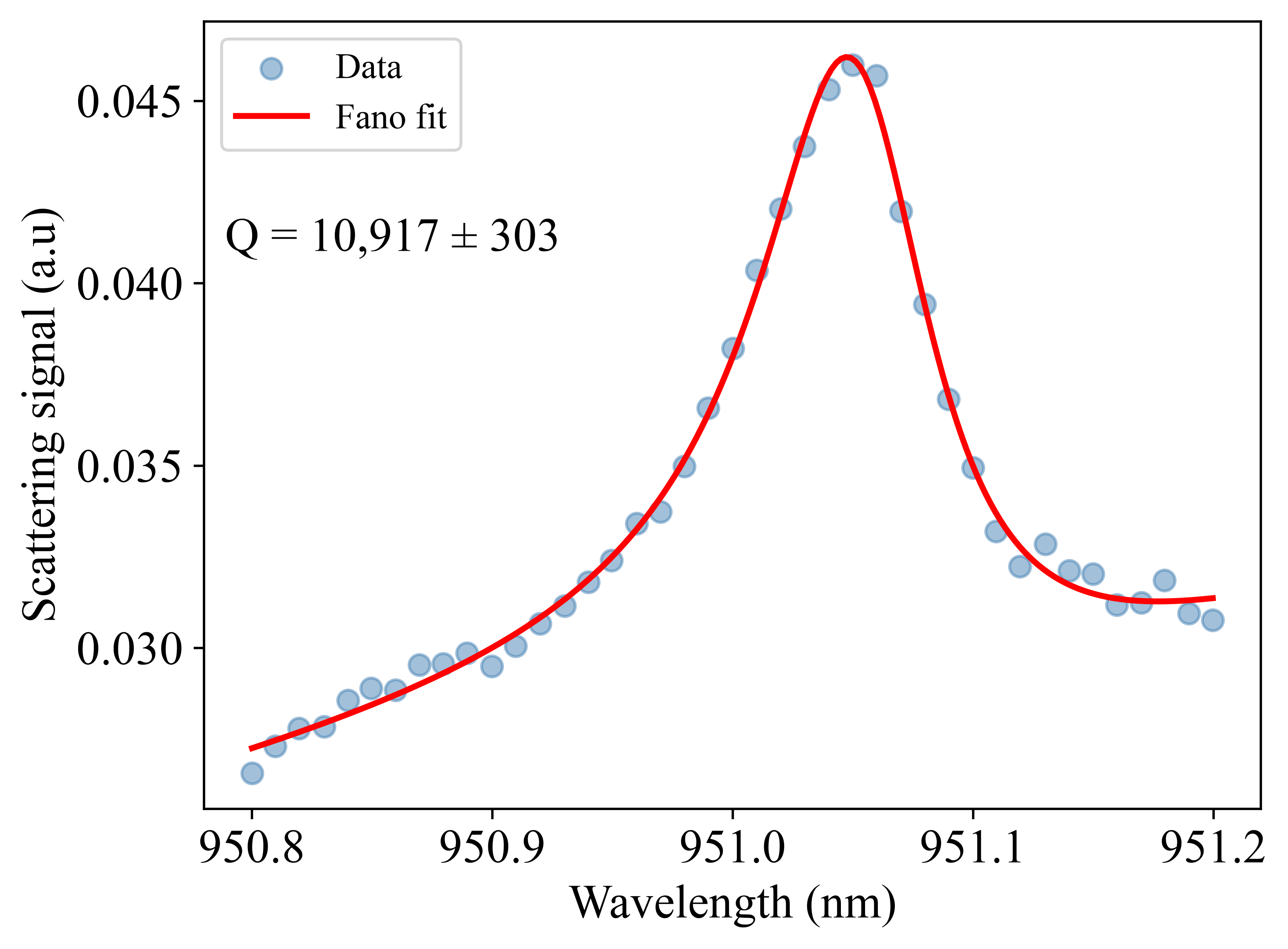}
        \caption{SC23107, device 17}
        \label{fig:device17}
    \end{subfigure}
    \\
    \begin{subfigure}[htbp]{0.495\linewidth}
        \centering
        \includegraphics[width=\linewidth]{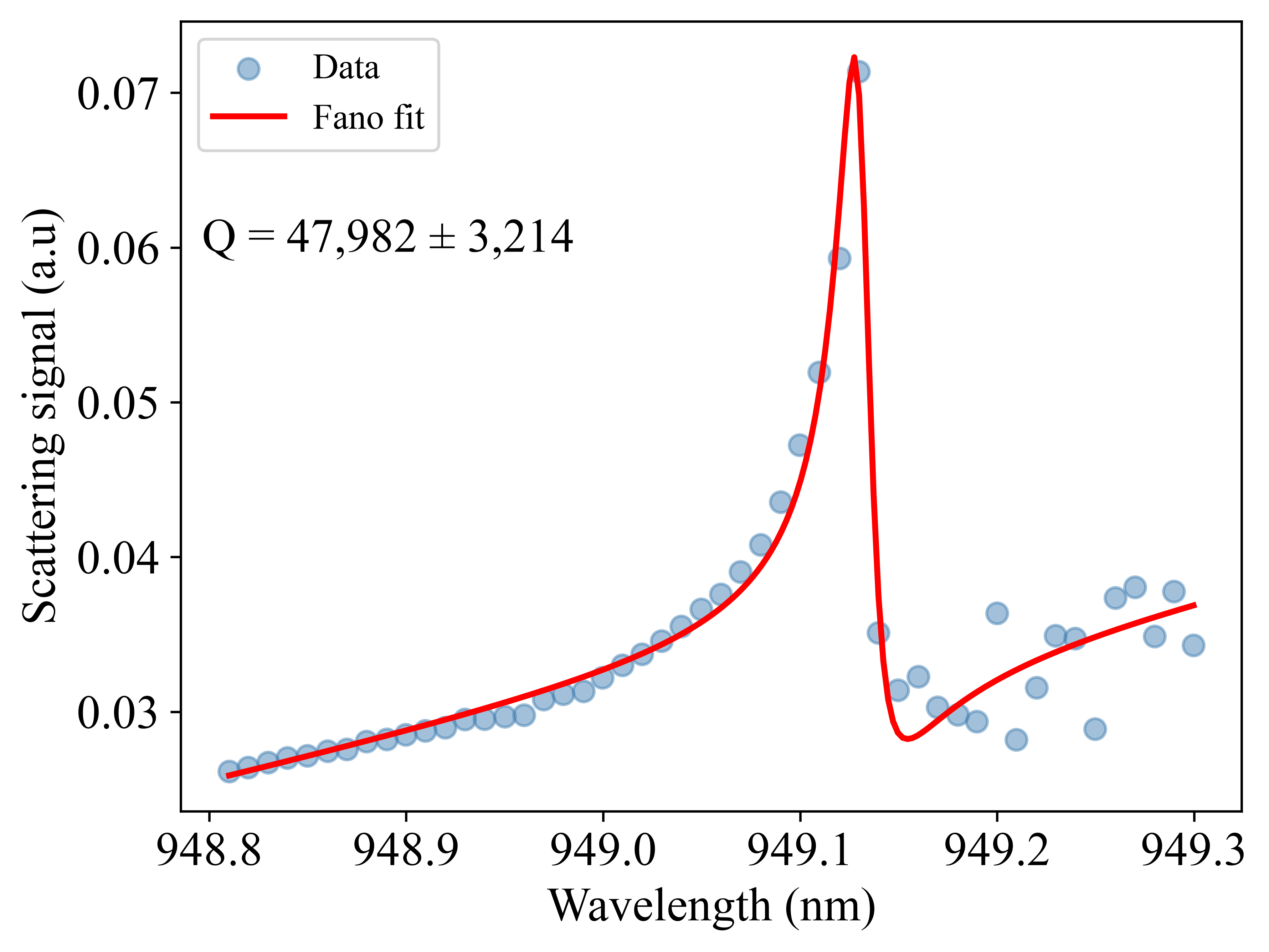} 
        \caption{SC23107, device 18}
        \label{fig:device18}
    \end{subfigure}
    \hfill
    \begin{subfigure}[htbp]{0.495\linewidth}
        \centering
        \includegraphics[width=\linewidth]{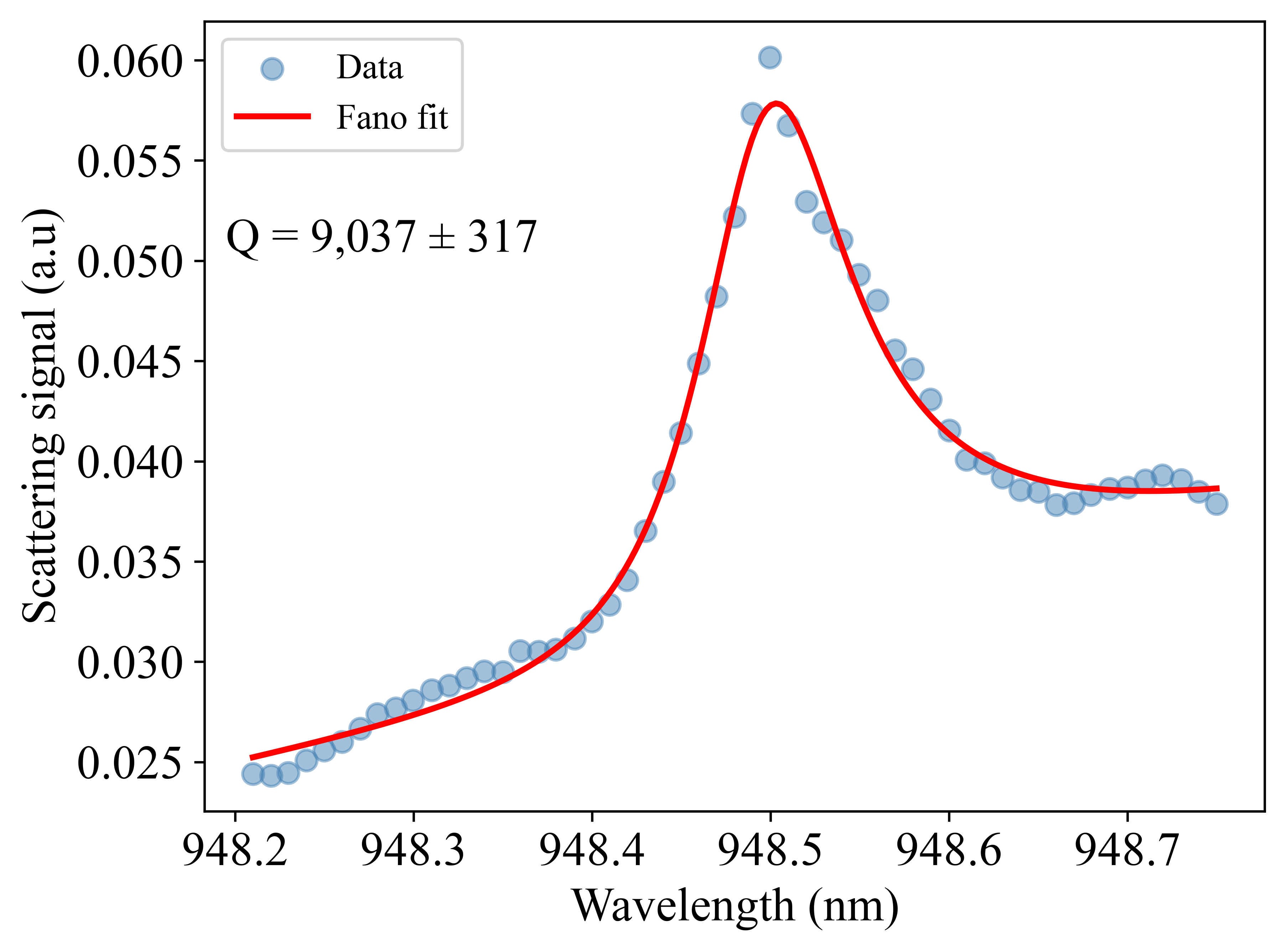}
        \caption{SC23107, device 21}
        \label{fig:device21}
    \end{subfigure}
    \caption{Cavity Q measurement of 12 devices with the same design as the device presented in the main text's Fig.~\ref{fig:device20}.}
    \label{fig:devices_summary}
\end{figure}

\printbibliography
\end{document}